\documentclass[a4paper,11pt]{article}
\pdfoutput=1 

\usepackage{jcappub} 

\usepackage{graphicx}
\usepackage{dcolumn}
\usepackage{bm}
\usepackage{physics}
\usepackage{xcolor}
\usepackage{comment}
\usepackage{amsmath}
\usepackage{mathrsfs}

\newcommand{\mpl}{M_{\mathrm{pl}}}
\newcommand{\bx}{\mathbf{x}}
\newcommand{\by}{\mathbf{y}}
\newcommand{\bp}{\mathbf{p}}
\newcommand{\bq}{\mathbf{q}}
\newcommand{\bk}{\mathbf{k}}
\newcommand{\mt}{\mathtt{t}}
\newcommand{\bz}{\bm{\zeta}}

\newcommand{\md}{\mathrm{d}}
\newcommand{\cb}{\mathscr{B}}

\title{\boldmath Comparing sharp and smooth transitions of the second slow-roll parameter in single-field inflation}

\author[1,2]{Jason Kristiano}
\author[3,1,2,4]{and Jun'ichi Yokoyama}

\affiliation[1]{Research Center for the Early Universe (RESCEU), Graduate School of Science, The University of Tokyo, Tokyo 113-0033, Japan}
\affiliation[2]{Department of Physics, Graduate School of Science, The University of Tokyo, Tokyo 113-0033, Japan}
\affiliation[3]{Kavli Institute for the Physics and Mathematics of the Universe (Kavli IPMU), WPI, UTIAS, The University of Tokyo, Kashiwa, Chiba 277-8568, Japan}
\affiliation[4]{Trans-Scale Quantum Science Institute, The University of Tokyo, Tokyo 113-0033, Japan}

\emailAdd{jkristiano@resceu.s.u-tokyo.ac.jp}
\emailAdd{yokoyama@resceu.s.u-tokyo.ac.jp}

\abstract{
In single-field inflation, violation of the slow-roll approximation can lead to growth of curvature perturbation outside the horizon. This violation is characterized by a period with a large negative value of the second slow-roll parameter. At an early time, inflation must satisfy the slow-roll approximation, so the large-scale curvature perturbation can explain the cosmic microwave background fluctuations. At intermediate time, it is viable to have a theory that violates the slow-roll approximation, which implies amplification of the curvature perturbation on small scales. Specifically, we consider ultraslow-roll inflation as the intermediate period. At late time, inflation should go back to the slow roll period so that it can end. This means that there are two transitions of the second slow-roll parameter. In this paper, we compare two different possibilities for the second transition: sharp and smooth transitions. Focusing on effects generated by the relevant cubic self-interaction of the curvature perturbation, we find that the bispectrum and one-loop correction to the power spectrum due to the change of the second slow-roll parameter vanish if and only if the
Mukhanov-Sasaki equation for perturbation satisfies a specific condition called Wands duality.
We also find in the case of sharp transition that, even though this duality is satisfied in the ultraslow-roll and slow-roll phases, it is severely violated at the transition so that the resultant one-loop correction is extremely large inversely proportional to the duration of the transition.
}

\begin{document}
\maketitle
\flushbottom

\section{Introduction}
We have recently shown that inflation models realizing a large-amplitude perturbation on small scales may suffer from excessive quantum corrections even on large scales probed by the cosmic microwave background (CMB) \cite{Kristiano:2022maq,Kristiano:2023scm, Kristiano:2024ngc}, adopting the same methodology as in our previous papers \cite{Kristiano:2021urj,Kristiano:2022zpn}. This observation has stimulated lively discussion in the community \cite{Riotto:2023hoz, Riotto:2023gpm, Choudhury:2023vuj, Choudhury:2023jlt, Choudhury:2023rks, Franciolini:2023lgy, Davies:2023hhn, Jackson:2023obv, Mishra:2023lhe, Choudhury:2024jlz, Firouzjahi:2023ahg, Iacconi:2023ggt, Motohashi:2023syh, Tasinato:2023ukp, Tasinato:2023ioq, Firouzjahi:2024psd, Maity:2023qzw, Firouzjahi:2023aum, Fumagalli:2023hpa, Tada:2023rgp, Firouzjahi:2023bkt, Cheng:2023ikq, Saburov:2024und, Ballesteros:2024zdp, Inomata:2024lud} because single-field inflation models realizing an appreciable abundance of primordial black holes (PBHs) \cite{Carr:2009jm, Carr:2020gox} are severely constrained.  

PBH formation can be well accommodated within the inflationary universe framework, which has been a part of standard cosmology to explain the early universe before the hot Big Bang \cite{Starobinsky:1980te,Sato:1980yn,Guth:1980zm} (reviewed in \cite{Sato:2015dga}). The canonical single-field slow-roll (SR) inflation is the most minimal theory to explain the observations of CMB anisotropy \cite{Planck:2018nkj, Planck:2018jri, Planck:2019kim}. It is a theory of a scalar field called inflaton with a canonical kinetic term and a potential term in a quasi-de Sitter (dS) background. At a certain region, the potential must be slightly tilted to realize SR dynamics of the inflaton, so the observed CMB fluctuations can be explained. Beyond that region, there are no significant observational constraints \cite{Nakama:2014vla, Jeong:2014gna, Inomata:2016uip, Nakama:2017qac, Kawasaki:2021yek, Kimura:2021sqz, Wang:2022nml}, so it is viable to have features in the potential that leads to violation of the SR condition, which is characterized by a large negative value of the second SR parameter (reviewed in \cite{Kristiano:2024ngc}). 

The simplest example of such a feature is a flat region in the potential \cite{Ivanov:1994pa}, which leads to the ultraslow-roll (USR) dynamics of the inflaton \cite{Kinney:1997ne, Inoue:2001zt, Kinney:2005vj, Martin:2012pe, Motohashi:2017kbs}. Moreover, it can be generalized to the constant-roll condition \cite{Motohashi:2014ppa, Motohashi:2017aob, Motohashi:2019rhu}.
During the SR violating period, the curvature perturbation can grow even at the superhorizon regime that yields amplification of the curvature perturbation on small scales \cite{Yokoyama:1998rw, Saito:2008em}. Such large fluctuations can collapse into primordial black holes after they re-enter the horizon during the radiation-dominated Universe \cite{Zel:1967, Hawking:1971ei, Carr:1974nx, Hawking:1974rv}. Currently, primordial black holes are proposed to be a dark matter candidate \cite{Chapline:1975ojl, Garcia-Bellido:1996mdl, Ivanov:1994pa, Yokoyama:1995ex, Afshordi:2003zb, Frampton:2010sw, Belotsky:2014kca, Carr:2016drx, Inomata:2017okj, Espinosa:2017sgp} (reviewed in \cite{Green:2020jor, Carr:2020xqk}) and explain both gravitational waves observed by LIGO-Virgo-KAGRA \cite{Sasaki:2016jop} and the pulsar timing array \cite{NANOGrav:2023gor, NANOGrav:2023hvm}, if they are formed with appreciable abundances. However, inflation must not end in such a non-slow-roll period. The inflaton must go back to a period in which the curvature perturbation does not grow outside the horizon. The simplest one is that the inflaton goes back to the SR period until the end of inflation. Thus, it implies that the existence of a transition from a temporary SR violating period to SR period is necessary.

We have claimed that the transition of the second SR parameter induces a large one-loop correction to the large-scale power spectrum probed by CMB \cite{Kristiano:2022maq}\footnote{One-loop correction to the power spectrum around peak scale has been discussed in \cite{Fumagalli:2023loc, Inomata:2022yte, Caravano:2024tlp}.}. Such a large backreaction casts doubt on the validity of perturbation theory even on a large scale. In order for the perturbative approach to be reliable, there is an upper bound on the small-scale power spectrum associated with the transition of the second SR parameter. Such a bound is also a requirement for performing order-by-order renormalization in perturbative theory. 

We have calculated one-loop corrections by assuming a sharp transition of the second SR parameter. However, our claim has been criticized by \cite{Riotto:2023hoz}. In \cite{Kristiano:2023scm}, we have responded by showing that the one-loop correction can be reproduced even by the method used in \cite{Riotto:2023hoz}. Then, our response has been further criticized by \cite{Riotto:2023gpm}, which argues that the one-loop correction cannot be generalized to a model with a smooth transition of the second SR parameter. By citing a specific case of smooth transition that yields a suppressed bispectrum \cite{Cai:2018dkf}, it is argued in \cite{Riotto:2023hoz, Riotto:2023gpm} that such a suppression should also be in the one-loop correction. In this paper, we analyze this claim more precisely to show that it is generically incorrect except for the particular case the Wands duality is satisfied as introduced below.

Apart from these criticisms, our works have been followed by many papers that mostly agree with our result, at least qualitatively. Renormalization, regularization, and resummation of the one-loop correction has been studied in \cite{Choudhury:2023vuj, Choudhury:2023jlt, Choudhury:2023rks}, where deviation from the canonical kinetic term is introduced as the sound speed of the curvature perturbation within the framework of the effective field theory of inflation. Our analytical formula for the one-loop correction has been confirmed by numerical integration in the sharp transition limit, starting by defining a time evolution of the second SR parameter \cite{Franciolini:2023lgy} or a generic inflationary potential \cite{Davies:2023hhn}. Moreover, there has been some progress toward the $\delta N$ formalism and stochastic inflation in the presence of a sharp transition of the second SR parameter \cite{Jackson:2023obv, Mishra:2023lhe, Choudhury:2024jlz}. The one-loop correction has been derived within the $\delta N$ formalism in \cite{Firouzjahi:2023ahg, Iacconi:2023ggt}. In \cite{Motohashi:2023syh}, the setup is extended to a general constant-roll period that leads to a constraint between the small-scale power spectrum and a constant second SR parameter. In extreme cases where the second SR parameter $\abs{\eta} \rightarrow \infty$ for a very short duration $\Delta \tau \rightarrow 0$ with finite $\abs{\eta} \Delta \tau$, bispectrum and one-loop correction have been investigated in \cite{Tasinato:2023ukp, Tasinato:2023ioq}. Moreover, loop correction to the bispectrum has been investigated in \cite{Firouzjahi:2024psd}.

In our papers, we have focused on the one-loop correction generated by the cubic Hamiltonian. In addition to such a contribution, another contribution from the quartic Hamiltonian is also expected, which has been derived either by Taylor expansion of the potential up to quartic order \cite{Maity:2023qzw} or an effective field theory approach \cite{Firouzjahi:2023aum}. Although it changes the precise value of the one-loop correction, one thing is for sure that it does not cancel the one-loop correction induced by cubic self-interaction. However, there are two papers that claim to find a vanishing one-loop correction \cite{Fumagalli:2023hpa, Tada:2023rgp}. In our work, we have implemented the field redefinition method to remove some terms in the cubic self-interaction which are either a total time derivative or a term proportional to the second-order equation of motion. In \cite{Fumagalli:2023hpa, Tada:2023rgp}, the one-loop correction is derived directly from the cubic Hamiltonian with total time derivative terms. Currently, they have been criticized by \cite{Firouzjahi:2023bkt}. We hope to solve this discrepancy in the future by pointing out what are missing in \cite{Fumagalli:2023hpa, Tada:2023rgp}. Other perspectives on the role of total time derivative terms can be found in \cite{Sou:2022nsd, Ning:2023ybc, Braglia:2024zsl, Kawaguchi:2024lsw}. 

Another issue is related to the gauge condition in which the one-loop computation is performed. We have calculated the one-loop correction in the comoving gauge, where perturbation is represented by the curvature perturbation. Some authors prefer to compute in the flat-slicing gauge, where perturbation is represented by the inflaton. Prior to our work, quantum correction to the inflaton perturbation has been investigated in \cite{Syu:2019uwx, Cheng:2021lif} and followed up in \cite{Cheng:2023ikq}, which is consistent with our claim. The effect of a smooth transition of the second SR parameter in the flat-slicing gauge has been explored in \cite{Saburov:2024und, Ballesteros:2024zdp}. However, there is a result \cite{Inomata:2024lud} that is inconsistent with our claim due to improper order counting.

The purpose of this paper is to compare the effect of sharp and smooth transitions of the second SR parameter with the correlation function of the curvature perturbation. For the sharp transition, the evolution of the second SR parameter is modeled as a step function of time. For a smooth transition, we will study a specific evolution that was proposed by \cite{Cai:2018dkf} and has been followed by many references afterwards. In 
Sect.~\ref{2pt}, we compare the differences in two-point functions or the power spectrum by solving the Mukhanov-Sasaki equation for both cases. In Sect.~\ref{3pt}, we compare the differences in three-point functions or bispectrum generated by the leading cubic self-interaction of curvature perturbation with the first-order expansion of in-in perturbation theory. In Sect.~\ref{1loop}, we compare the differences in one-loop correction to the power spectrum generated by the same cubic self-interaction of curvature perturbation with the second-order expansion of in-in perturbation theory. Finally, Sect.~\ref{con} is devoted to the conclusion.  

\section{Two-point functions \label{2pt}}
Consider a scalar field $\phi$ as the inflaton. The action of canonical inflation is given by
\begin{equation}
S = \frac{1}{2} \int \md^4x \sqrt{-g} \left[ \mpl^2 R - (\partial_\mu \phi)^2 - 2 V(\phi) \right], \label{action}
\end{equation}
where $\mpl$ is reduced Planck scale, $g = \mathrm{det}~g_{\mu\nu}$, $g_{\mu\nu}$ and $R$ are metric tensor and its Ricci scalar. The homogeneous part of the inflaton, $\phi(t)$, evolves on the spatially flat, homogeneous, and isotropic background,
\begin{equation}
\md s^2 = -\md t^2 + a^2(t) \md \mathbf{x}^2 = a^2(\tau) (-\md \tau^2 + \md \mathbf{x}^2),
\end{equation}
where $t$ is physical time and $\tau$ is conformal time. Equations of motion for the scale factor $a(t)$ and $\phi(t)$ are the Friedmann equations
\begin{equation}
H^2 = \frac{1}{3\mpl^2}\left(\frac{1}{2} \dot{\phi}^2 + V(\phi)\right), ~\dot{H} = - \frac{\dot{\phi}^2}{2 \mpl^2}, \label{fried}
\end{equation}
with $H=\dot{a}/{a}$ being the Hubble parameter, and the Klein-Gordon equation
\begin{equation}
\ddot{\phi} + 3 H \dot{\phi} + \frac{\md V}{\md\phi} = 0. \label{klein}
\end{equation}
In this paper, a dot denotes the derivative with respect to $t$.

During inflation, the evolution of the Hubble parameter is parameterized by the slow-roll (SR) parameters. The first SR parameter $\epsilon$ is defined as
\begin{equation}
\epsilon \equiv - \frac{\dot{H}}{H^2} = \frac{\dot{\phi}^2}{2 \mpl^2 H^2} , \label{epst}
\end{equation}
where the equality is obtained by substituting Eq. \eqref{fried}. When the Hubble parameter is almost constant, or $\epsilon \ll 1$, the scale factor can be approximated as
\begin{equation}
a \approx - \frac{1}{H \tau} \propto e^{H t},
\end{equation}
which is called quasi-de Sitter (dS) approximation. The second SR parameter, which parameterizes evolution of the first SR parameter, is defined as
\begin{equation}
\eta \equiv \frac{\dot{\epsilon}}{\epsilon H} =  2 \frac{\ddot{\phi}}{\dot{\phi} H} + 2 \epsilon .
\end{equation}
When the inequalities
\begin{equation}
\epsilon, \abs{\eta}, \abs{\frac{\dot{\eta}}{\eta H}} \ll 1,
\end{equation}
are satisfied, we say the SR approximation is fulfilled.  
Therefore, SR approximation implies quasi-dS solution. However, the quasi-dS approximation does not imply SR approximation, because the second SR parameter can have a large negative value while keeping $\epsilon \ll 1$.

Small perturbation from the homogeneous part, $\phi(t)$, of the inflaton $\phi(\mathbf{x}, t)$ and metric can be expressed as 
\begin{gather}
\phi(\mathbf{x},t) = \phi(t) + \delta \phi(\mathbf{x},t), \nonumber\\
\md s^2 = -N^2 \mathrm{d}t^2 + \gamma_{ij} (\mathrm{d}x^i + N^i \mathrm{d}t)(\mathrm{d}x^j + N^j \mathrm{d}t),
\end{gather}
where $\gamma_{ij}$ is the three-dimensional metric on slices of constant $t$, $N$ is the lapse function, and $N^i$ is the shift vector. We choose comoving gauge condition
\begin{equation}
\delta \phi(\mathbf{x},t) = 0, ~\gamma_{ij}(\mathbf{x},t) = a^2(t) e^{2\zeta(\mathbf{x},t)} \delta_{ij},
\end{equation}
where $\zeta(\mathbf{x},t)$ is curvature perturbation. $N$ and $N^i$ are obtained by solving the constraint equations. In this paper, we do not discuss the tensor perturbation.

The second-order expansion of \eqref{action} is given by
\begin{equation}
S^{(2)}[\zeta] = M_{\mathrm{pl}}^2 \int \mathrm{d}t ~\mathrm{d}^3x ~a^3 \epsilon  \left[ \dot{\zeta}^2 - \frac{1}{a^2} (\partial_i \zeta)^2  \right].
\label{S2}
\end{equation}
In terms of Mukhanov-Sasaki variable $v = z \mpl \zeta$, where $z = a \sqrt{2\epsilon}$, the action becomes canonically normalized
\begin{equation}
S^{(2)}_{\mathrm{can}} [v] = \frac{1}{2} \int \mathrm{d}\tau ~\mathrm{d}^3x \left[ (v')^2 - (\partial_i v)^2 + \frac{z''}{z} v^2 \right],
\end{equation}
where a prime denotes derivative with respect to $\tau$. We can read $z''/z$ as an effective time-dependent mass squared of the Mukhanov-Sasaki variable. In momentum space, quantization is performed by expanding the Mukhanov-Sasaki variable in terms of the creation $\hat{a}^\dagger_\bk$ and the annihilation $\hat{a}_\bk$ operator,
\begin{equation}
v_\bk (\tau) = M_{\mathrm{pl}} z(\tau) \zeta_\bk (\tau) =  v_k(\tau) \hat{a}_{\mathbf{k}} + v^*_k (\tau) \hat{a}_{-\mathbf{k}}^\dagger, \nonumber
\end{equation}
where mode function $v_k(\tau)$ satisfies the Mukhanov-Sasaki equation
\begin{equation}
v_k'' + \left( k^2 - \frac{z''}{z} \right) v_k = 0.
\end{equation}
The operators satisfy the commutation relation $[ \hat{a}_{\mathbf{k}}, \hat{a}_{-\mathbf{k'}}^\dagger ] = (2 \pi)^3 \delta(\mathbf{k} + \mathbf{k'})$ under the normalization condition
\begin{equation}
v_k'^* v_k - v_k' v_k^* = i. \label{normalization}
\end{equation}
In terms of the SR parameters, the effective mass can be written as
\begin{equation}
\frac{z''}{z} = (a H)^2 \left( 2 - \epsilon +\frac{3}{2}\eta -\frac{1}{2} \epsilon \eta + \frac{1}{4} \eta^2 + \frac{\dot{\eta}}{2 H} \right).
\end{equation}
Note that this expression is exact, and we have not performed any approximation yet.

In SR approximation, the effective mass can be approximated as
\begin{equation}
\frac{z''}{z} \simeq \frac{2}{\tau^2} .
\end{equation}
The general solution of mode function $v_k(\tau)$ is
\begin{equation}
v_k(\tau) = \frac{\mathcal{A}_k}{\sqrt{2k}} \left( 1-\frac{i}{k\tau} \right) e^{-ik\tau} + \frac{\mathcal{B}_k}{\sqrt{2k}} \left( 1+\frac{i}{k\tau} \right) e^{ik\tau}, \label{mssol}
\end{equation}
where $\mathcal{A}_k$ and $\mathcal{B}_k$ are determined by boundary conditions or definition of a vacuum state. Imposing the Bunch-Davies vacuum at $\tau \rightarrow - \infty$, defined by $\hat{a} \ket{0} = 0$, implies $\mathcal{A}_{k} = 1$ and $\mathcal{B}_{k} = 0$. Then the mode function of the curvature perturbation $\zeta_k = v_k/ z \mpl$ is given by
\begin{equation}
\zeta_k(\tau) = \left( \frac{i H}{2 \mpl \sqrt{k^3 \epsilon}} \right)_\star (1+ik\tau) e^{-ik\tau}, \label{zetads}
\end{equation}
subscript $\star$ denotes the value at the horizon crossing epoch $\tau = -1/k$.

The two-point function of curvature perturbation and power spectrum can be written as
\begin{gather}
\langle \zeta_\bk (\tau) \zeta_{\bk'} (\tau) \rangle \equiv (2 \pi)^3 \delta(\mathbf{k} + \mathbf{k'}) \langle \! \langle \zeta_\bk (\tau) \zeta_{-\bk} (\tau) \rangle \! \rangle, \\
\Delta^2_s(k, \tau) \equiv \frac{k^3}{2 \pi^2} \langle \! \langle \zeta_\bk (\tau) \zeta_{-\bk} (\tau) \rangle \! \rangle, 
\end{gather}
where the bracket $\langle \cdots \rangle = \bra{0} \cdots \ket{0}$ denotes the vacuum expectation value (VEV), and $\Delta^2_s(k,\tau)$ is the power spectrum multiplied by the phase space density. At late time, $ \tau = \tau_0 (\rightarrow 0)$, the power spectrum approaches an almost scale-invariant limit
\begin{equation}
\Delta^2_{s (\mathrm{SR}) }(k) \equiv \Delta^2_{s }(k, \tau_0) = \left( \frac{H^2}{8 \pi^2 M_{\mathrm{pl}}^2 \epsilon} \right)_\star, \label{pssr}
\end{equation}
with a weak wavenumber dependence due to the horizon crossing condition manifested in the spectral tilt
\begin{equation}
n_s(k, \tau_0) - 1 = \frac{\md \log \Delta_s^{2}(k, \tau_0)}{\md \log k} = (- 2 \epsilon - \eta)_\star.
\end{equation}

Observation of the power spectrum at wavenumber $p$ provides information about the first SR parameter $\epsilon$ evaluated at $\tau = - 1/p$. CMB observations tightly constrain the power spectrum at the wavenumber $0.005 \mathrm{Mpc}^{-1} < p < 0.2 \mathrm{Mpc}^{-1}$. At the pivot scale, $p_* = 0.05 ~\mathrm{Mpc}^{-1}$, the power spectrum is $\Delta_{s(\mathrm{CMB})}^{2} \equiv \Delta^2_{s (\mathrm{SR}) }(p_*) = 2.1 \times 10^{-9}$. This means that the first SR parameter is only tightly constrained within a specific period corresponding to the CMB horizon crossing time at around $60$ e-folds before the end of inflation.

After such an epoch, if the first SR parameter decreases rapidly, the power spectrum can be amplified on small scales $k \gg p$. This can be achieved by violating the SR approximation at a later time, specifically $\eta \sim \mathcal{O}(-1)$. Taking into account that $\eta$ can evolve in general, the effective mass becomes
\begin{equation}
\frac{z''}{z} = (a H)^2 \left[ 2 +\frac{3}{2}\eta + \frac{1}{4} \eta^2 + \frac{\dot{\eta}}{2H}  + \mathcal{O}(\epsilon) \right].
\end{equation}

The simplest example of a violation of the SR approximation is USR inflation \cite{Kinney:1997ne,Inoue:2001zt, Kinney:2005vj, Martin:2012pe, Motohashi:2017kbs}. This happens when the inflaton passes through an extremely flat region of the potential. In this region, the first derivative of the potential vanishes, $\md V / \md \phi \approx 0$, so the Klein-Gordon equation \eqref{klein} becomes $\ddot{\phi} \approx -3H \dot{\phi}$. It leads to $\dot{\phi} \propto a^{-3}$, so the SR parameters become
\begin{equation}
\epsilon = \frac{\dot{\phi}^2}{2 \mpl^2 H^2}  \propto a^{-6} \Longrightarrow \eta \approx -6, \label{epsusr}
\end{equation}
which shows that SR approximation is broken.

To terminate inflation, after a period when the SR approximation is violated, the inflaton must return to an attractor period. In this period, the curvature perturbation must not grow outside the horizon and its power spectrum must not increase further on smaller scales. A SR period certainly satisfies such requirements. More generally, a period with constant second SR parameter, $\eta \gtrsim 0$, called constant roll, is also feasible.

\begin{figure}[tbp]
\centering 
\includegraphics[width=0.5\textwidth]{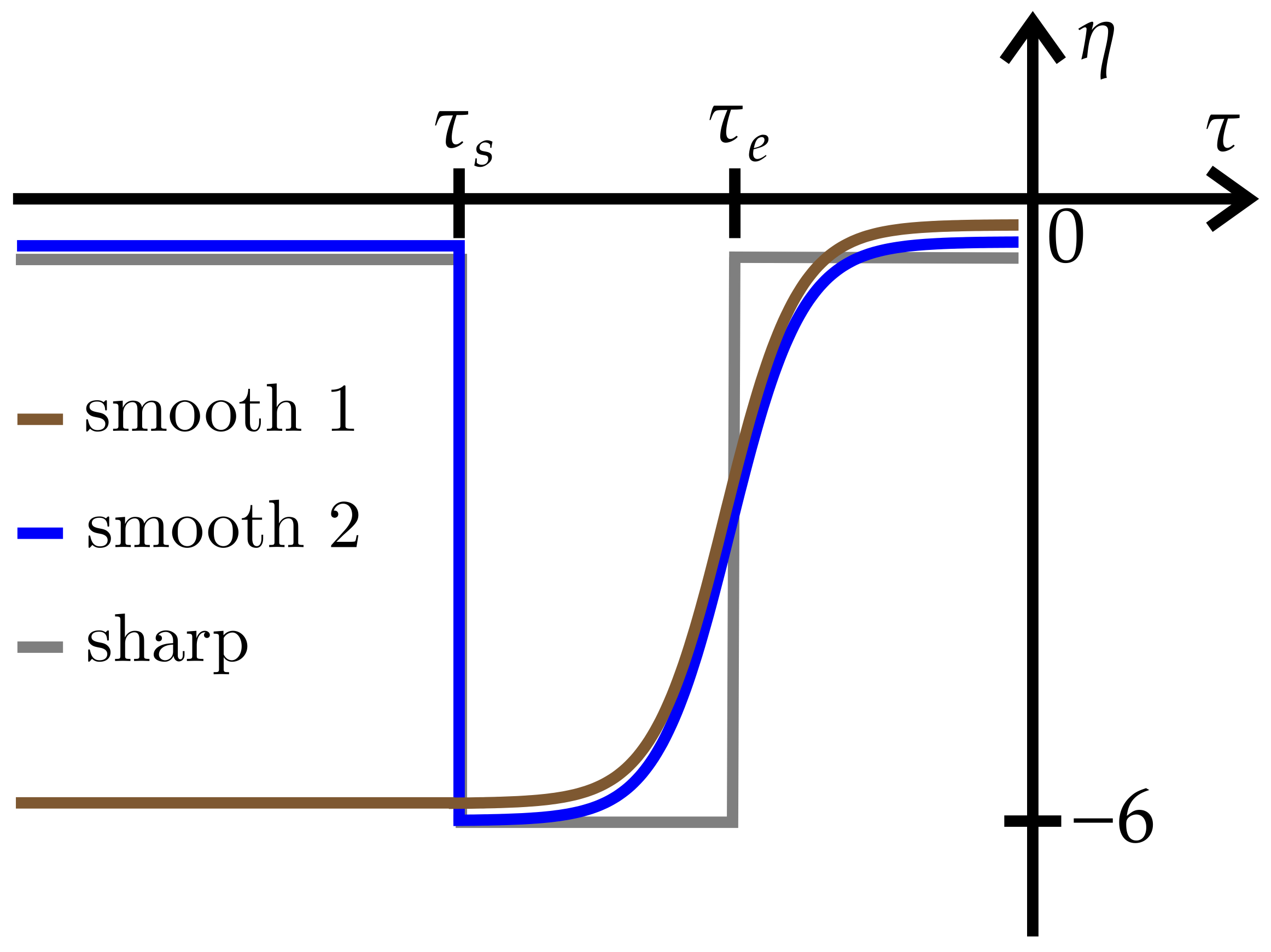}
\caption{\label{fig1} Sketch of sharp and smooth transition of second SR parameter.}
\end{figure}

Hence, in a viable model of single-field inflation relizing PBH formation, the inflaton undergoes two transitions between the CMB horizon crossing time until the end of inflation, one from SR to USR and the other from USR to SR.
In the following subsections, we want to compare the effect of sharp and smooth second transition to the power spectrum, specifically the gray and blue curves in Fig.~\ref{fig1}.

\subsection{Sharp transition}
Consider evolution of the second SR parameter, as shown by the gray curve in Fig.~\ref{fig1}. The second SR parameter reads as
\begin{equation}
\eta(\tau) = \begin{cases}
0 & ; \tau \leq \tau_s \\
-6 & ; \tau_s < \tau \leq \tau_e \\
0 & ; \tau > \tau_e
\end{cases} ~.
\end{equation}
Integrating it yields the first SR parameter
\begin{equation}
\epsilon(\tau) = \begin{cases}
\epsilon(\tau_s) & ; \tau \leq \tau_s \\
\epsilon(\tau_s) \left( \dfrac{\tau}{\tau_s} \right)^6 & ; \tau_s < \tau \leq \tau_e \\
\epsilon(\tau_s) \left( \dfrac{\tau_e}{\tau_s} \right)^6 & ; \tau > \tau_e
\end{cases} ~,
\end{equation}
where $\epsilon(\tau_s)$ is taken as a reference. At any time except around the transitions, the effective mass is $z''/z \simeq 2/\tau^2$.

\begin{figure}[tbp]
\centering 
\includegraphics[width=0.75\textwidth]{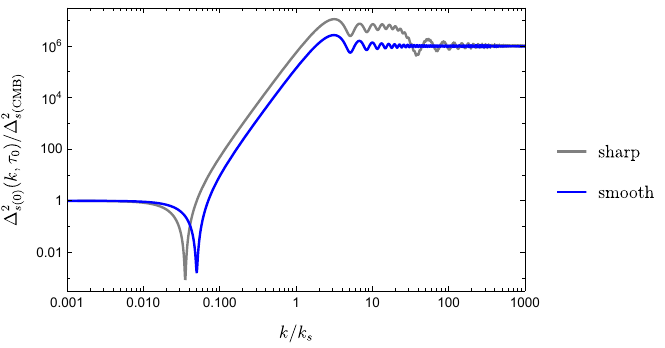}
\caption{\label{fig5} Comparing power spectrum between the sharp and smooth transition case.}
\end{figure}

The mode function of curvature perturbation is
\begin{equation}
\zeta_k(\tau) = \begin{cases}
\dfrac{i H}{2 \mpl \sqrt{ k^3\epsilon(\tau_s)}}  (1+ik\tau) e^{-ik\tau} & ; \tau \leq \tau_s \\
\dfrac{i H}{2 \mpl \sqrt{ k^3\epsilon(\tau_s)}} \left( \dfrac{\tau_s}{\tau} \right)^3 \left[ \mathcal{A}_k e^{-ik\tau} (1+ik\tau) - \mathcal{B}_k e^{ik\tau} (1-ik\tau) \right] & ; \tau_s < \tau \leq \tau_e \\
\dfrac{i H}{2 \mpl \sqrt{ k^3 \epsilon(\tau_s)}} \left( \dfrac{\tau_s}{\tau_e} \right)^3  \left[ \mathcal{C}_k e^{-ik\tau} (1+ik\tau) - \mathcal{D}_k e^{ik\tau} (1-ik\tau) \right] & ; \tau > \tau_e
\end{cases} ~, \label{zetasr2}
\end{equation}
where coefficients $\mathcal{A}_k$, $\mathcal{B}_k$, $\mathcal{C}_k$, and $\mathcal{D}_k$ are to be determined from boundary conditions \cite{Starobinsky:1992ts, Leach:2001zf, Byrnes:2018txb, Liu:2020oqe, Tasinato:2020vdk, Karam:2022nym, Pi:2022zxs}. We neglect the horizon crossing condition in the mode function because it is not important for our discussion. Coefficients $\mathcal{A}_k$ and $\mathcal{B}_k$ are obtained by requiring continuity of $\zeta_k(\tau)$ and $\zeta_k'(\tau)$ at transition $\tau = \tau_s$
\begin{gather}
\mathcal{A}_k = 1 - \frac{3(1 + k^2 \tau_s^2)}{2i k^3 \tau_s^3} \label{coefa2}, \\
\mathcal{B}_k = - \frac{3(1 + i k \tau_s)^2}{2i k^3 \tau_s^3} e^{-2ik \tau_s}, \label{coefb2}
\end{gather}
while coefficients $\mathcal{C}_k$ and $\mathcal{D}_k$ are obtained by requiring the same conditions at transition $\tau = \tau_e$
\begin{align}
\mathcal{C}_k =  \frac{-1}{4k^6 \tau_s^3 \tau_e^3} & \left\lbrace 9(k\tau_s - i)^2 (k\tau_e + i)^2 e^{2ik(\tau_e - \tau_s)} \right. \nonumber\\
& \left. - \left[k^2 \tau_s^2 (2 k \tau_s + 3i) + 3i \right] \left[k^2 \tau_e^2 (2k \tau_e - 3i) - 3i \right] \right\rbrace, \label{coefa3}
\end{align}
\begin{align}
\mathcal{D}_k = \frac{3}{4k^6 \tau_s^3 \tau_e^3} & \left\lbrace e^{-2i k \tau_s} [3 + k^2 \tau_e^2 (3-2i k \tau_e)] (k \tau_s - i)^2 \right. \nonumber\\
& \left. + i e^{-2i k \tau_e} \left[ 3i + k^2 \tau_s^2 (2 k \tau_s + 3i) \right] (k \tau_e - i)^2 \right\rbrace. \label{coefb3}
\end{align}
Thus, power spectrum at the end of inflation is 
\begin{equation}
\Delta^2_{s(0)}(k, \tau_0) = \frac{k^3}{2 \pi^2} \abs{\zeta_k(\tau_0)}^2 =  \frac{H^2}{8 \pi^2 M_{\mathrm{pl}}^2 \epsilon(\tau_s)}  \left( \frac{k_e}{k_s} \right)^6 \abs{\mathcal{C}_k - \mathcal{D}_k}^2, \label{ps0}
\end{equation}
where subscript $(0)$ denotes tree-level contribution, $k_s$ and $k_e$ are wavenumber crossing the horizon at $\tau_s$ and $\tau_e$, respectively.
\begin{equation}
\Delta_{s(\mathrm{PBH})}^{2} \approx \Delta_{s(\mathrm{SR})}^{2}(k_s) \left( \frac{k_e}{k_s} \right)^6.
\end{equation}
Plot of the typical power spectrum is shown in Fig.~\ref{fig5}.

\subsection{Smooth transition}
Consider evolution of the second SR parameter, as shown by the blue curve in Fig.~\ref{fig1}. In contrast to the previous subsection, in this case $\eta(t)$ evolves smoothly from USR period back to SR period. We adopt a parametrization that is used in many literatures \cite{Cai:2018dkf, Riotto:2023gpm, Riotto:2023hoz, Firouzjahi:2023ahg, Iacconi:2023ggt},
\begin{equation}
\frac{3}{2}\eta + \frac{1}{4}\eta^2 + \frac{\dot{\eta}}{2 H} = \mathrm{constant} \equiv \frac{3}{2}\delta. \label{etadif}
\end{equation}
With this parametrization, the effective mass becomes
\begin{equation}
\frac{z''}{z} = \frac{1}{\tau^2} \left( 2 + \frac{3}{2} \delta \right) \equiv \frac{1}{\tau^2} \left( \nu^2 - \frac{1}{4} \right),
\end{equation}
where a parameter $\nu$ is defined. For $\delta \rightarrow 0$, solution of the Mukhanov-Sasaki equation is given by \eqref{mssol}. Introducing $s\equiv \sqrt{9 + 6\delta} \simeq 3 + \delta$, the differential equation for $\eta(\tau)$ is
\begin{equation}
\int_{-6}^\eta \frac{\md \eta}{(\eta + 3)^2 - s^2} = \int_{\tau_s}^\tau \frac{\md \tau}{2 \tau} ,
\end{equation}
and integrating it yields a solution
\begin{equation}
\eta(\tau) = s-3 - \frac{2s (s+3)}{ \left( \dfrac{\tau_s}{\tau} \right)^s (s-3) + s + 3 } . \label{etasmooth}
\end{equation}
It shows that the second SR parameter evolves from the USR period $\eta(\tau_s) = - 6$ to the SR period $\eta(\tau_0) = \delta$, so $\delta$ is the second SR parameter at the end of inflation. Next, we can obtain the time derivative of $\eta(\tau)$ as
\begin{equation}
\eta'(\tau) = \frac{-2 s^2(s-3)(s+3)  }{\tau \left[ \left( \dfrac{\tau_s}{\tau} \right)^s (s-3) + s + 3 \right]^2} \left( \frac{\tau_s}{\tau} \right)^s,
\end{equation}
and find the time derivative at initial time is
\begin{equation}
\eta'(\tau_s) = - \frac{1}{2 \tau_s} (s-3)(s+3) \simeq - \frac{3 }{\tau_s} \delta.
\end{equation}

Taking a time derivative of the differential equation \eqref{etadif} yields
\begin{equation}
0 = 2 \eta' + \eta' \eta - \eta'' \tau ,
\end{equation}
which in turn means  $\epsilon(\tau) a^2(\tau) \eta'(\tau) = \mathrm{constant}$, because we find 
\begin{equation}
(\epsilon a^2 \eta' )' = \epsilon a^3 H \left( 2\eta' + \eta \eta' -\eta'' \tau \right) = 0.
\end{equation}
 Therefore, $\epsilon(\tau)$ can be expressed as
\begin{equation}
\epsilon(\tau) = \epsilon(\tau_s) \frac{a^2(\tau_s) \eta'(\tau_s)}{a^2(\tau) \eta'(\tau)} = \epsilon(\tau_s) \left[ \frac{s + 3}{2s} \left( \frac{\tau}{\tau_s} \right)^{\frac{3+s}{2}} + \frac{s - 3}{2s} \left( \frac{\tau}{\tau_s} \right)^{\frac{3-s}{2}} \right]^2 .
\end{equation}
In terms of e-fold number $N \equiv \log (\tau_s/\tau)$, it can be written as
\begin{equation}
\epsilon(N) = \epsilon(0) \left( \frac{s + 3}{2s} e^{- \frac{3+s}{2} N} + \frac{s - 3}{2s} e^{\frac{s-3}{2} N} \right)^2.
\end{equation}
Then, evolution of the Hubble parameter is given by
\begin{align}
\log \frac{H(N)}{H(0)} = \frac{1}{4s^2} \epsilon(0) & \left[ (s+3) \left( e^{-(s+3)N} - 1 \right) + (s-3) \left(1 - e^{(s-3)N} \right) \right. \nonumber\\
&\left. + \frac{2}{3}(s+3) (s-3) \left( e^{-3N} - 1 \right) \right].
\end{align}

From \eqref{epst}, we can obtain background evolution of the inflaton as
\begin{align}
\Delta\phi(N) & \equiv  \phi(N) - \phi(0) = \mpl \int_{0}^N \md N \sqrt{2 \epsilon(N)} \nonumber\\
& = \frac{\mpl}{s} \sqrt{2 \epsilon(0)} \left( e^{\frac{s-3}{2}N} - e^{-\frac{s+3}{2}N} \right).
\end{align}
Since $s \cong 3+\delta$ for small $\delta$, the first term in the right hand side will
soon dominate over the second for $N \gtrsim 1$, so that we can reverse the equation to yield
\begin{equation}
e^{N} \cong \left( \frac{3}{\sqrt{2 \epsilon(0)}} \frac{\Delta\phi}{\mpl} \right)^{\frac{2}{\delta}} ,
\end{equation}
and the potential $V = \mpl^2 H^2 (3 - \epsilon)$ reads
\begin{align}
V(\Delta\phi) \cong & \mpl^2 H^2(0) \left\lbrace 3 - \epsilon(0)  \left[ \left( \frac{3}{\sqrt{2 \epsilon(0)}} \frac{\Delta\phi}{\mpl}  \right)^{-\frac{6}{\delta}} + \frac{\delta}{2}  \frac{1}{\sqrt{2 \epsilon(0)}} \frac{\Delta\phi}{\mpl}   \right]^2   \right\rbrace \nonumber\\
\times \exp & \left\lbrace \frac{1}{3} \epsilon(0) \left[ \left( \frac{3}{\sqrt{2 \epsilon(0)}} \frac{\Delta\phi}{\mpl}  \right)^{-\frac{12}{\delta}} - 1 \right] + \frac{\delta}{18} \epsilon(0) \left[ 1 - \left( \frac{3}{\sqrt{2 \epsilon(0)}} \frac{\Delta\phi}{\mpl} \right)^2 \right] \right. \nonumber\\
&\left. + \frac{2 \delta}{9} \epsilon(0)  \left[ \left( \frac{3}{\sqrt{2 \epsilon(0)}} \frac{\Delta\phi}{\mpl} \right)^{-\frac{6}{\delta}} - 1 \right] \right\rbrace.
\end{align}

\begin{figure}[tbp]
\centering 
\includegraphics[width=0.75\textwidth]{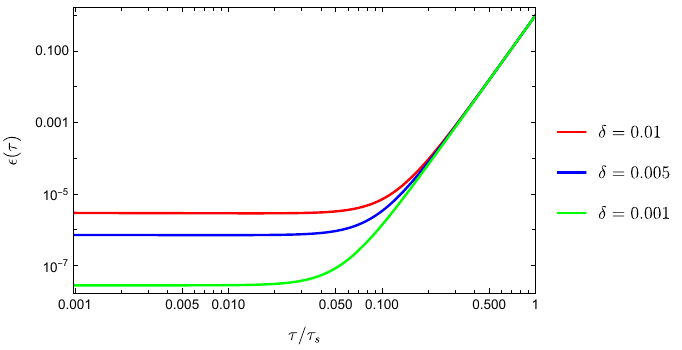}
\caption{\label{fig4} Plot of $\epsilon(\tau)$ for various $\delta$.}
\end{figure}

\begin{figure}[tbp]
\centering 
\includegraphics[width=0.75\textwidth]{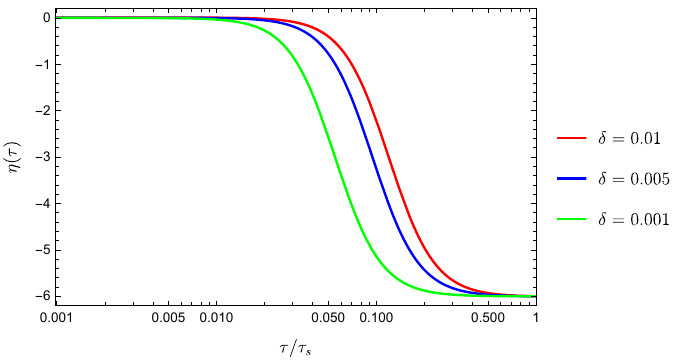}
\caption{\label{fig3} Plot of $\eta(\tau)$ for various $\delta$.}
\end{figure}

\begin{figure}[tbp]
\centering 
\includegraphics[width=0.75\textwidth]{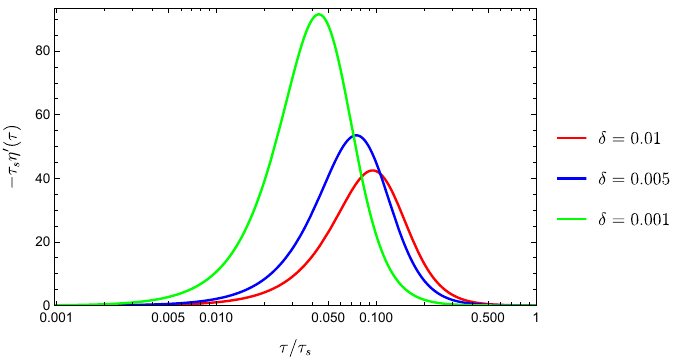}
\caption{\label{fig2} Plot of $\eta'(\tau)$ for various $\delta$.}
\end{figure}

The plots of $\epsilon(\tau)$, $\eta(\tau)$ and $\eta'(\tau)$ are shown in Figs. \ref{fig4}, \ref{fig3}, and \ref{fig2}, respectively. We can see that $\eta'(\tau)$ has a maximum point that is located at
\begin{equation}
\left( \frac{\tau_m}{\tau_s} \right)^{-3-\delta} = \frac{(4 + \delta) (6 + \delta)}{\delta (2 + \delta)} \Longrightarrow \tau_m \simeq \tau_s \left( \frac{\delta}{12} \right)^{1/3} ,
\end{equation}
which corresponds to
\begin{equation}
\eta'(\tau_m) \simeq - \frac{4}{\tau_s} \left( \frac{12}{\delta} \right)^{1/3}.
\end{equation}
Interestingly, the second SR parameter when its time derivative reaches its maximum value is $\eta(\tau_m) = -2$. As a remark, we can see that describing evolution \eqref{etasmooth} as a smooth transition is misleading, because the maximum value of $\eta'(\tau)$ or the smoothness is controlled by $\delta$. For $\delta \rightarrow 0$, $\eta'(\tau)$ can be very large, which implies that the transition is sharp. Thus, we prefer to call evolution \eqref{etasmooth} as a transition that satisfies Wands duality condition \cite{Wands:1998yp}. It is a duality that relates two different background conditions with the same Mukhanov-Sasaki equation. For a constant $\eta$, the simplest example of Wands duality is USR with $\eta= -6$ and SR with $\eta = 0$, which have the same effective mass $z''/z = 2/\tau^2$. For a dynamical $\eta$, Wands duality condition \eqref{etadif} implies that the curvature perturbation satisfies the same Mukhanov-Sasaki equation at any time during the transition.

Then, we would like to find a mode function of the curvature perturbation in this setup. Requiring continuity of $\zeta_k(\tau)$ and $\zeta_k'(\tau)$ at $\tau = \tau_s$ corresponds to (dis)continuity of $v_k(\tau)$ and $v_k'(\tau)$ that reads
\begin{gather}
v_k(\tau_s^-) = v_k(\tau_s^+) , \\
v_k'(\tau_s^+) - v_k'(\tau_s^-) = v_k (\tau_s) \Delta\left( \frac{z'}{z} \right) = v_k(\tau_s) \frac{\Delta\eta}{2} aH .
\end{gather}
We rewrite the effective mass as
\begin{equation}
\frac{z''}{z} = \begin{cases}
\dfrac{1}{\tau^2} \left( \nu_1^2 - \dfrac{1}{4} \right)  & ;\tau < \tau_s \\
\dfrac{1}{\tau^2} \left( \nu_2^2 - \dfrac{1}{4} \right)  & ;\tau \geq \tau_s
\end{cases} ,
\end{equation}
where $\nu_1 = 3/2$ and $\nu_2 = \delta + 3/2$.
We define dimensionless time $\mt \equiv - k \tau$, so the transition time can be denoted as $\mt_s \equiv - k \tau_s$. Solution of the Mukhanov-Sasaki equation with Bunch-Davies initial condition is 
\begin{equation}
v_k(\mt) = \begin{cases}
\sqrt{ \dfrac{\pi \mt}{4k} } e^{i (2\nu_1 + 1) \frac{\pi}{4} } H_{\nu_1}^{(1)}(\mt) & ;\mt > \mt_s \\
\sqrt{\mt} \left[ \alpha_k H_{\nu_2}^{(1)}(\mt) + \beta_k H_{\nu_2}^{(2)}(\mt) \right] & ;\mt \leq \mt_s
\end{cases} ~~,
\end{equation}
where $H_{\nu}^{(1)}(\mt)$ and $H_{\nu}^{(2)}(\mt)$ are the Hankel function of the first and second kind, respectively. Its time derivative is
\begin{equation}
\frac{\md}{\md \mt} v_k(\mt) = \begin{cases}
\sqrt{ \dfrac{\pi}{4 k \mt} } e^{i (2\nu_1 + 1) \frac{\pi}{4} } \left[ \left( \dfrac{1}{2} - \nu_1 \right) H_{\nu_1}^{(1)}(\mt) + \mt H_{\nu_1 - 1 }^{(1)}(\mt) \right] & ;\mt > \mt_s \\
\dfrac{\alpha_k}{\sqrt{\mt}}  \left[ \left( \dfrac{1}{2} - \nu_2 \right) H_{\nu_2}^{(1)}(\mt) + \mt H_{\nu_2 - 1 }^{(1)}(\mt) \right]  \nonumber\\
 + \dfrac{\beta_k}{\sqrt{\mt}}  \left[ \left( \dfrac{1}{2} - \nu_2 \right) H_{\nu_2}^{(2)}(\mt) + \mt H_{\nu_2 - 1 }^{(2)}(\mt) \right] & ;\mt \leq \mt_s
\end{cases} ~~.
\end{equation}

\begin{figure}[tbp]
\centering 
\includegraphics[width=0.75\textwidth]{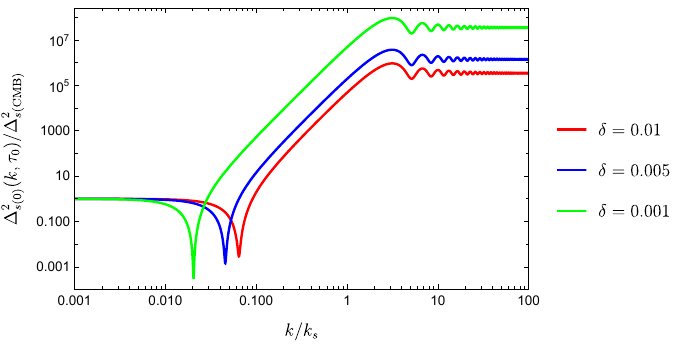}
\caption{\label{fig6} Power spectrum for various $\delta$ in the smooth transition case.}
\end{figure}

Coefficients $\alpha_k$ and $\beta_k$ are determined by requiring (dis)continuity of $v_k(\tau)$ and $v_k'(\tau)$ at $\tau = \tau_s$
\begin{gather}
\alpha_k H_{\nu_2}^{(1)}(\mt_s) + \beta_k H_{\nu_2}^{(2)}(\mt_s) = \frac{\pi}{4k} e^{i (2\nu_1 + 1) \frac{\pi}{4} } H_{\nu_1}^{(1)}(\mt_s), \\
\alpha_k \left[ \left( \frac{1}{2} - \nu_2 + \frac{k_s}{2k} \Delta\eta \mt_s \right) H_{\nu_2}^{(1)}(\mt_s) + \mt_s H_{\nu_2 - 1 }^{(1)}(\mt_s) \right] + \beta_k \left[ \left( \frac{1}{2} - \nu_2 + \frac{k_s}{2k} \Delta\eta \mt_s \right) H_{\nu_2}^{(2)}(\mt_s) \right. \nonumber\\
\left. + \mt_s H_{\nu_2 - 1 }^{(2)}(\mt_s) \right] = \sqrt{\frac{\pi}{4k}} e^{i (2\nu_1 + 1) \frac{\pi}{4} } \left[ \left( \frac{1}{2} - \nu_1 \right) H_{\nu_1}^{(1)}(\mt_s) + \mt_s H_{\nu_1 - 1 }^{(1)}(\mt_s) \right] .
\end{gather}
The (dis)continuity equations can be understood as linear equations
\begin{align}
a_1 \alpha_k + b_1 \beta_k & = c_1 ,\nonumber\\
a_2 \alpha_k + b_2 \beta_k & = c_2 ,
\end{align}
with coefficients \cite{Mishra:2023lhe}
\begin{align}
a_1 & = H_{\nu_2}^{(1)}(\mt_s) , \nonumber\\
b_1 & = H_{\nu_2}^{(2)}(\mt_s) , \nonumber\\
c_1 & =  \sqrt{\frac{\pi}{4k}} e^{i (2\nu_1 + 1) \frac{\pi}{4} } H_{\nu_1}^{(1)}(\mt_s) , \nonumber\\
a_2 & = \left( \frac{1}{2} - \nu_2 + \frac{k_s}{2k} \Delta\eta \mt_s \right) H_{\nu_2}^{(1)}(\mt_s) + \mt_s H_{\nu_2 - 1 }^{(1)}(\mt_s) , \nonumber\\
b_2 & = \left( \frac{1}{2} - \nu_2 + \frac{k_s}{2k} \Delta\eta \mt_s \right) H_{\nu_2}^{(2)}(\mt_s) + \mt_s H_{\nu_2 - 1 }^{(2)}(\mt_s) , \nonumber\\
c_2 & = \sqrt{\frac{\pi}{4k}} e^{i (2\nu_1 + 1) \frac{\pi}{4} }  \left[ \left( \frac{1}{2} - \nu_1 \right) H_{\nu_1}^{(1)}(\mt_s) + \mt_s H_{\nu_1 - 1 }^{(1)}(\mt_s) \right].
\end{align}
The solutions of the linear equations are
\begin{equation}
\alpha_k = \frac{c_1 b_2 - c_2 b_1}{a_1 b_2 - a_2 b_1} ~\mathrm{and}~ \beta_k = \frac{c_2 a_1 - c_1 a_2}{a_1 b_2 - a_2 b_1} ,
\end{equation}
where the denominator can be further simplified as $a_1 b_2 - a_2 b_1 = -4 i / \pi$. Thus, the curvature perturbation after the transition is
\begin{equation}
\zeta_k(\tau) = \frac{H}{\mpl \sqrt{2 \epsilon(\tau)}} (-\tau) \sqrt{-k \tau} \left[ \alpha_k H_{\nu_2}^{(1)}(-k \tau) + \beta_k H_{\nu_2}^{(2)}( -k \tau) \right], \label{modesmooth}
\end{equation}
with power spectrum
\begin{equation}
\Delta^2_{s(0)}(k, \tau_0) = \frac{k^3}{2 \pi^2} \abs{\zeta_k(\tau \rightarrow 0)}^2.
\end{equation}
The plot of the power spectrum with various $\delta$ is shown in Fig.~\ref{fig6}. Compared to the CMB-scale power spectrum, the small-scale power spectrum is amplified by
\begin{equation}
\Delta_{s(\mathrm{PBH})}^{2} \approx \Delta_{s(\mathrm{SR})}^{2}(k_s) \left( \frac{3}{\delta} \right)^2.
\end{equation}

\begin{figure}[tbp]
\centering 
\includegraphics[width=0.75\textwidth]{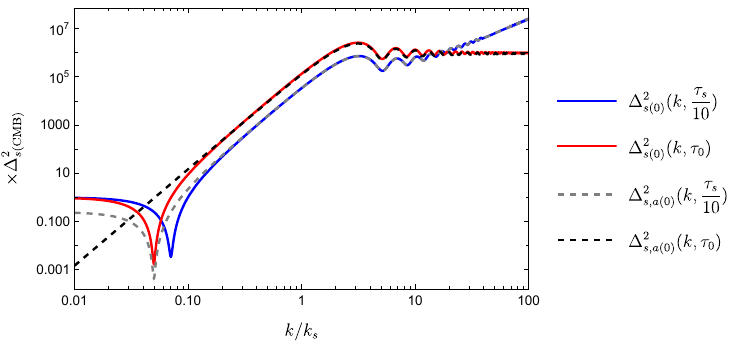}
\caption{\label{fig9} Comparing the exact and approximate formula of the power spectrum for various $\delta$ in the smooth transition case at two different times: $\tau_s/10$ and $\tau_0 \rightarrow 0$.}
\end{figure}

However, the exact formula for the curvature perturbation is complicated due to the Hankel function with index $\nu_2 = \delta + 3/2$. One might wonder what happens if $\nu_2 = 3/2$ is substituted. In that case, the curvature perturbation becomes
\begin{equation}
\zeta^a_k(\tau) = \frac{i H}{2 \mpl \sqrt{ k^3 \epsilon(\tau)}} \left[ \mathcal{A}_k e^{-ik\tau} (1+ik\tau) - \mathcal{B}_k e^{ik\tau} (1-ik\tau) \right], 
\end{equation}
with power spectrum
\begin{equation}
\Delta^2_{s,a(0)}(k, \tau) \equiv \frac{k^3}{2 \pi^2} \langle \! \langle \zeta_\bk^a (\tau) \zeta_{-\bk}^a (\tau) \rangle \! \rangle = \frac{H^2}{8 \pi^2 \mpl^2 \epsilon(\tau)} \abs{\mathcal{A}_k - \mathcal{B}_k}^2,
\end{equation}
where $\mathcal{A}_k$ and $\mathcal{B}_k$ are given by \eqref{coefa2} and \eqref{coefb2}, respectively. In Fig.~\ref{fig9}, we can see that the approximation works well only on small scales, but cannot reproduce the correct behavior on large scales. This happens because the Hankel function has an important dependence on $\delta$ when the argument approaches zero. Therefore, to obtain the correct prediction from small to large scale, the precise mode function \eqref{modesmooth} must be used.

\section{Three-point functions \label{3pt}}
In the previous section, we have shown that sharp and smooth transitions of the second SR parameter predict almost the same power spectrum. Then, one might ask whether the transition can be distinguished by higher-point functions. In this section, we investigate the difference between the three-point functions generated by sharp and smooth transitions.

Third-order expansion of \eqref{action} is given by \cite{Maldacena:2002vr}
\begin{equation}
S^{(3)}[\zeta] = S_{\mathrm{bulk}}[\zeta] + S_\mathrm{B}[\zeta] + \mpl^2  \int \md t ~\md^3 x ~2 f(\zeta) \left( \frac{\delta L}{\delta \zeta} \right)_1. \label{S3}
\end{equation}
The bulk interaction $S_{\mathrm{bulk}}[\zeta]$ reads 
\begin{align}
S_{\mathrm{bulk}}[\zeta] = \mpl^2  \int \md t ~\md^3 x ~a^3 & \left[ \epsilon^2  \dot{\zeta}^2 \zeta + \frac{1}{a^2} \epsilon^2 (\partial_i \zeta)^2 \zeta - 2 \epsilon  \dot{\zeta} \partial_i \zeta \partial_i \chi \right. \nonumber\\
& \left. -\frac{1}{2} \epsilon^3 \dot{\zeta}^2 \zeta + \frac{1}{2} \epsilon \zeta (\partial_i \partial_j \chi)^2 + \frac{1}{2} \epsilon \dot{\eta} \dot{\zeta} \zeta^2 \right], \label{s3bulk}
\end{align}
where $\chi = \epsilon \partial^{-2} \dot{\zeta}$. 
$S_\mathrm{B}[\zeta]$ contains total time derivative terms that are explicitly given in \cite{Arroja:2011yj, Burrage:2011hd}. The last term is interaction proportional to the linear equation of motion.
\begin{equation}
\left( \frac{\delta L}{\delta \zeta} \right)_1 = \partial_t (\epsilon a^3 \dot{\zeta}) - \epsilon a \partial^2 \zeta,
\end{equation}
and function $f(\zeta)$ is given by
\begin{equation}
f(\zeta) = \frac{\eta}{4} \zeta^2 + \frac{\dot{\zeta}}{H} \zeta + \frac{1}{4a^2 H^2} [-(\partial_i \zeta)^2 + \partial^{-2}\partial_i \partial_j (\partial_i \zeta \partial_j \zeta)] + \frac{1}{2H}[\partial_i \zeta \partial_i \chi - \partial^{-2} \partial_i \partial_j (\partial_i \zeta \partial_j \chi)]. \label{redef}
\end{equation}
Then, one can redefine the curvature perturbation so that its third-order action contains only the bulk interaction. Performing field redefinition $\zeta = \bz + f(\bz)$ so the sum of the second-order and third-order action becomes
\cite{Maldacena:2002vr, Arroja:2011yj, Burrage:2011hd}
\begin{equation}
S^{(2)}[\zeta] + S^{(3)}[\zeta] = S^{(2)}[\bz] + S_{\mathrm{bulk}}[\bz]. \label{sbz}
\end{equation}

Next, we want to calculate the three-point functions generated by a transition of the second SR parameter. In this case, the relevant interaction is the last term in \eqref{s3bulk} \cite{Namjoo:2012aa, Cai:2016ngx, Chen:2013eea, Cai:2018dkf, Passaglia:2018ixg, Davies:2021loj}. Also, at the end of inflation, the field redefinition is $\zeta = \bz$ because the inflaton is in SR period with $\eta \approx 0$. Three-point functions is calculated from the first-order in-in perturbation theory
\begin{equation}
\langle \mathcal{O(\tau)} \rangle = 2 \int_{-\infty}^{\tau} \md\tau_1 ~\mathrm{Im} \left\langle \mathcal{\hat{O}} (\tau) H^{(3)}(\tau_1) \right\rangle, \label{inin1}
\end{equation}
where operator $\mathcal{O(\tau)}$ is $\zeta_{\bk_1}(\tau) \zeta_{\bk_2}(\tau) \zeta_{\bk_3}(\tau)$. The Hamiltonian can be obtained from the third-order action with relation $H^{(3)} = - \int \md^3x ~\mathcal{L}_3$, where $\mathcal{L}_3$ is the integrand of \eqref{s3bulk}. The relevant Hamiltonian, which corresponds to the last term in \eqref{s3bulk} reads 
\begin{equation}
H^{(3)}(\tau) = - \frac{1}{2} \mpl^2 \int \md^3 x ~\epsilon(\tau) \eta'(\tau) a^2(\tau) \zeta'(\bx,\tau) \zeta^2(\bx,\tau). \label{hint}
\end{equation}
Strictly speaking, it should be the Hamiltonian of $\bz$. With it, one can calculate the three-point functions of $\bz$. Because the three-point functions of $\zeta$ and $\bz$ are equal at the end of inflation, we can use $\zeta$ in the calculation to simplify the notation\footnote{We are aware that there is a subtlety in the equivalence of deriving correlation function at second-order in-in perturbation theory between direct implementation of total time derivative terms $S_\mathrm{B}[\zeta]$ and field redefinition \cite{Fumagalli:2023hpa, Tada:2023rgp}. However, this issue is beyond the scope of this paper. The purpose of this paper is to compare the effect of the sharp and smooth transition of $\eta(\tau)$ to the correlation functions generated by the last term in \eqref{s3bulk}.}.

Bispectrum $\langle \! \langle \zeta_{\bk_1}(\tau) \zeta_{\bk_2}(\tau) \zeta_{\bk_3}(\tau) \rangle \!\rangle$ is defined as
\begin{equation}
\langle \zeta_{\bk_1}(\tau) \zeta_{\bk_2}(\tau) \zeta_{\bk_3}(\tau) \rangle = (2\pi)^3 \delta(\bk_1 + \bk_2 + \bk_3) \langle \! \langle \zeta_{\bk_1}(\tau) \zeta_{\bk_2}(\tau) \zeta_{\bk_3}(\tau) \rangle \! \rangle.
\end{equation}
Substituting \eqref{hint} to \eqref{inin1}, we obtain \cite{Kristiano:2023scm}
\begin{align}
\langle\!\langle \zeta_{\bk_1}(\tau_0) \zeta_{\bk_2}(\tau_0) \zeta_{\bk_3}(\tau_0) \rangle\!\rangle = & - 2\mpl^2 \int_{-\infty}^{\tau_0} \md \tau ~\epsilon(\tau) \eta'(\tau) a^2(\tau) \\
& \times \mathrm{Im} \left[ \zeta_{k_1}(\tau_0) \zeta_{k_2}(\tau_0) \zeta_{k_3}(\tau_0) \zeta_{k_1}^*(\tau) \zeta_{k_2}^*(\tau) \zeta_{k_3}'^*(\tau) + \mathrm{perm} \right]. \nonumber
\end{align}
There are two independent contributions to the bispectrum, reads
\begin{equation}
\langle\!\langle \zeta_{\bk_1}(\tau_0) \zeta_{\bk_2}(\tau_0) \zeta_{\bk_3}(\tau_0) \rangle\!\rangle = \langle\!\langle \zeta_{\bk_1}(\tau_0) \zeta_{\bk_2}(\tau_0) \zeta_{\bk_3}(\tau_0) \rangle\!\rangle^\mathrm{(s)} + \langle\!\langle \zeta_{\bk_1}(\tau_0) \zeta_{\bk_2}(\tau_0) \zeta_{\bk_3}(\tau_0) \rangle\!\rangle^\mathrm{(e)}.
\end{equation}
The superscript (s) denotes the contribution from the sharp transition from the SR to the USR period at $\tau = \tau_s$. During this transition, $\eta'(\tau)$ can be approximated by $\eta'(\tau) = - \Delta\eta ~\delta(\tau-\tau_s)$. Thus, this transition contributes to the bispectrum \cite{Kristiano:2022maq}
\begin{align}
\langle\!\langle \zeta_{\bk_1}(\tau_0) \zeta_{\bk_2}(\tau_0) \zeta_{\bk_3}(\tau_0) \rangle\!\rangle^\mathrm{(s)} = & ~2\mpl^2 \epsilon(\tau_s) a^2(\tau_s)  \Delta\eta \\
& \times \mathrm{Im} \left[ \zeta_{k_1}(\tau_0) \zeta_{k_2}(\tau_0) \zeta_{k_3}(\tau_0) \zeta_{k_1}^*(\tau_s) \zeta_{k_2}^*(\tau_s) \zeta_{k_3}'^*(\tau_s) + \mathrm{perm} \right]. \nonumber
\end{align}
Superscript (e) denotes the transition from USR to final SR period. This transition can be sharp or smooth. In the following, we investigate both cases and compare the differences.

\subsection{Sharp transition}
\begin{figure}[tbp]
\centering 
\includegraphics[width=0.75\textwidth]{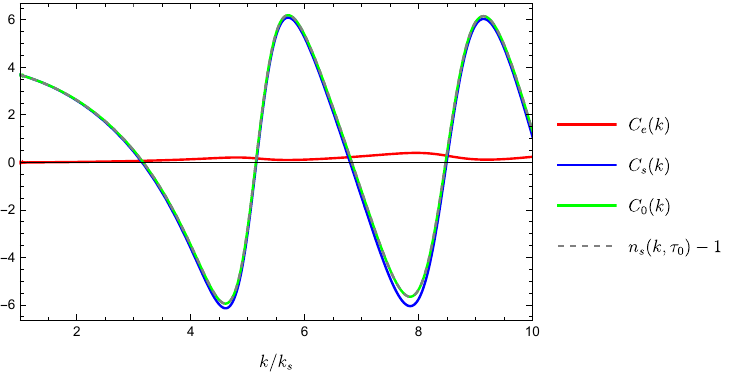}
\caption{\label{fig7} Plot of $C_s(k)$, $C_e(k)$, $C_0(k)$, and $n_s(k,\tau_0) - 1$ for the sharp transition case. We choose $k_e/k_s = 10$ just for illustrative purposes.}
\end{figure}

In this subsection, we consider a case where the transition from USR to the final SR period is also sharp. During this transition, $\eta'(\tau)$ can be approximated by $\eta'(\tau) = \Delta\eta ~\delta(\tau-\tau_e)$. Thus, this transition contributes to the bispectrum \cite{Kristiano:2023scm}
\begin{align}
\langle\!\langle \zeta_{\bk_1}(\tau_0) \zeta_{\bk_2}(\tau_0) \zeta_{\bk_3}(\tau_0) \rangle\!\rangle^\mathrm{(e)} = & -2\mpl^2 \epsilon(\tau_e) a^2(\tau_e)  \Delta\eta \\
& \times \mathrm{Im} \left[ \zeta_{k_1}(\tau_0) \zeta_{k_2}(\tau_0) \zeta_{k_3}(\tau_0) \zeta_{k_1}^*(\tau_e) \zeta_{k_2}^*(\tau_e) \zeta_{k_3}'^*(\tau_e) + \mathrm{perm} \right]. \nonumber
\end{align}
Total contributions to the bispectrum reads
\begin{gather}
\langle \! \langle \zeta_{\bk_1}(\tau_0) \zeta_{\bk_2}(\tau_0) \zeta_{-\bk_2}(\tau_0) \rangle \!\rangle = - \left\lbrace 4 \Delta\eta \mpl^2 \epsilon(\tau_e) a^2(\tau_e) \mathrm{Im} \left[ \frac{\zeta_{k_2}^2(\tau_0)}{\abs{\zeta_{k_2}(\tau_0)}^2} \zeta_{k_2}^*(\tau_e) \zeta_{k_2}'^*(\tau_e) \right] \right. \nonumber\\ 
 \left. - 4 \Delta\eta \mpl^2 \epsilon(\tau_s) a^2(\tau_s) \mathrm{Im} \left[ \frac{\zeta_{k_2}^2(\tau_0)}{\abs{\zeta_{k_2}(\tau_0)}^2} \zeta_{k_2}^*(\tau_s) \zeta_{k_2}'^*(\tau_s) \right] \right\rbrace \abs{\zeta_{k_1}(\tau_0)}^2 \abs{\zeta_{k_2}(\tau_0)}^2 .
\end{gather}
We define the first and second terms inside parenthesis as $C_e(k)$ and $C_s(k)$, respectively. We also denote the sum of $C_s(k)$ and $C_e(k)$ as $C_0(k)$. The plot of $C_s(k)$, $C_e(k)$, and $C_0(k)$ are shown in Fig. \ref{fig7}. We also show that the bispectrum satisfies Maldacena's theorem, namely $C_0(k) = n_s(k,\tau_0) - 1$ \cite{Kristiano:2023scm}. Interestingly, we find that $C_s(k)$ almost fits $n_s(k,\tau_0) - 1$. This means that the first transition contributes to the bispectrum much more than the second transition. 

\subsection{Smooth transition}
In this subsection, we consider the case where the second SR parameter evolves as a continuous function from the USR to the final SR period, assuming that the evolution of the second SR parameter for $\tau > \tau_s$ satisfies the differential equation \eqref{etadif}. Its contribution to the bispectrum is given by
\begin{align}
\langle\!\langle \zeta_{\bk_1}(\tau_0) \zeta_{\bk_2}(\tau_0) \zeta_{\bk_3}(\tau_0) \rangle\!\rangle^\mathrm{(e)} = & - 2\mpl^2 \int_{\tau_s}^{\tau_0} \md \tau ~\epsilon(\tau) \eta'(\tau) a^2(\tau) \label{tintsmooth}\\
& \times \mathrm{Im} \left[ \zeta_{k_1}(\tau_0) \zeta_{k_2}(\tau_0) \zeta_{k_3}(\tau_0) \zeta_{k_1}^*(\tau) \zeta_{k_2}^*(\tau) \zeta_{k_3}'^*(\tau) + \mathrm{perm} \right]. \nonumber
\end{align}
Using the fact that $\epsilon(\tau) a^2(\tau) \eta'(\tau)$ is a constant within the integration domain, the bispectrum can be simplified as
\begin{align}
\langle\!\langle \zeta_{\bk_1}(\tau_0) \zeta_{\bk_2}(\tau_0) \zeta_{\bk_3}(\tau_0) \rangle\!\rangle^\mathrm{(e)} = & - 2\mpl^2 \epsilon(\tau_s) \eta'(\tau_s) a^2(\tau_s) \label{smoothint} \\
& \times \mathrm{Im} \left\lbrace \zeta_{k_1}(\tau_0) \zeta_{k_2}(\tau_0) \zeta_{k_3}(\tau_0) \int_{\tau_s}^{\tau_0} \md \tau  \left[ \zeta_{k_1}^*(\tau) \zeta_{k_2}^*(\tau) \zeta_{k_3}^*(\tau) \right]' \right\rbrace. \nonumber
\end{align}
Performing the time integral yields
\begin{align}
\langle\!\langle \zeta_{\bk_1}(\tau_0) \zeta_{\bk_2}(\tau_0) \zeta_{\bk_3}(\tau_0) \rangle\!\rangle^\mathrm{(e)} = & ~2\mpl^2 \epsilon(\tau_s) \eta'(\tau_s) a^2(\tau_s) \label{smoothbe} \\
& \times \mathrm{Im} \left[ \zeta_{k_1}(\tau_0) \zeta_{k_2}(\tau_0) \zeta_{k_3}(\tau_0) \zeta_{k_1}^*(\tau_s) \zeta_{k_2}^*(\tau_s) \zeta_{k_3}^*(\tau_s) \right]. \nonumber
\end{align}

\subsubsection{Ultraslow-roll to slow-roll transition}
The purpose of this part is to compare our result with \cite{Cai:2018dkf}. Consider the evolution of the second SR parameter as shown by the brown curve in Fig.~\ref{fig1}. The second SR parameter reads as
\begin{equation}
\eta(\tau) = \begin{cases}
-6 & ; \tau \leq \tau_s \\
s-3 - \dfrac{2s (s+3)}{ \left( \dfrac{\tau_s}{\tau} \right)^s (s-3) + s + 3 } & ; \tau > \tau_s
\end{cases} ~.
\end{equation}  
In this setup, the long-wavelength curvature perturbation leaves the horizon at USR period. Also, the Bunch-Davies initial vacuum condition is imposed in the USR period unlike in our previous study \cite{Kristiano:2022maq,Kristiano:2023scm} and as we did in Sect.~\ref{2pt}.

The curvature perturbation is given by
\begin{equation}
\zeta_k(\tau) = \frac{i H}{2 \mpl \sqrt{ k^3 \epsilon(\tau)}}  e^{-ik\tau} (1+ik\tau). \label{modeusrsr}
\end{equation}
Substituting this mode function to \eqref{smoothbe} leads to
\begin{align}
\langle\!\langle \zeta_{\bk_1}(\tau_0) \zeta_{\bk_2}(\tau_0) \zeta_{\bk_3}(\tau_0) \rangle\!\rangle = & ~2\mpl^2 \epsilon(\tau_s) \eta'(\tau_s) a^2(\tau_s) \left( \frac{H^2}{4 \mpl^2 k_1 k_2 k_3 \sqrt{\epsilon(\tau_0) \epsilon(\tau_s) }} \right)^3  \\
& \times \mathrm{Im} \left[ e^{-i(k_1 + k_2 + k_3)/k_s} \left( 1 + \frac{i k_1}{k_s} \right) \left( 1 + \frac{i k_2}{k_s} \right) \left( 1 + \frac{i k_3}{k_s} \right) \right]. \nonumber
\end{align}
After some algebras, it becomes
\begin{align}
\langle\!\langle \zeta_{\bk_1}(\tau_0) \zeta_{\bk_2}(\tau_0) \zeta_{\bk_3}(\tau_0) \rangle\!\rangle = & ~2\mpl^2 \epsilon(\tau_s) \eta'(\tau_s) a^2(\tau_s) \left( \frac{H^2}{4 \mpl^2 k_1 k_2 k_3 \sqrt{\epsilon(\tau_0) \epsilon(\tau_s) }} \right)^3  \label{bsmoothfull}\\
& \times \frac{1}{k_s^3} \left[ ( -k_1 k_2 k_3 + (k_1 + k_2 + k_3) k_s^2 ) \cos
\frac{k_1 + k_2 + k_3}{k_s} \right. \nonumber\\
& \left. + k_s (k_1 k_2 + k_2 k_3 + k_1 k_3 - k_s^2) \sin \frac{k_1 + k_2 + k_3}{k_s}\right] . \nonumber
\end{align}
In squeezed limit, $k_1 \rightarrow 0$, the bispectrum becomes
\begin{align}
\langle\!\langle \zeta_{\bk_1}(\tau_0) \zeta_{\bk_2}(\tau_0) \zeta_{-\bk_2}(\tau_0) \rangle\!\rangle = &~ 2\mpl^2 \epsilon(\tau_s) \eta'(\tau_s) a^2(\tau_s) \left( \frac{H^2}{4 \mpl^2 k_1 k_2^2 \sqrt{\epsilon(\tau_0) \epsilon(\tau_s) }} \right)^3  \\
& \times \left[ \frac{2k_2}{k_s} \cos \frac{2k_2}{k_s} + \left( \frac{k_2^2}{k_s^2} - 1 \right) \sin \frac{2k_2}{k_s}\right] . \nonumber
\end{align}
For $k_2 \ll k_s$, which is considered in \cite{Cai:2018dkf}, the bispectrum becomes
\begin{align}
\langle\!\langle \zeta_{\bk_1}(\tau_0) \zeta_{\bk_2}(\tau_0) \zeta_{-\bk_2}(\tau_0) \rangle\!\rangle & = 2\mpl^2 \epsilon(\tau_s) \eta'(\tau_s) a^2(\tau_s) \left( \frac{H^2}{4 \mpl^2 k_1 k_2^2 \sqrt{\epsilon(\tau_0) \epsilon(\tau_s) }} \right)^3 \frac{2}{3} \left( \frac{k_2}{k_s} \right)^3 \nonumber\\
& = \delta \sqrt{\frac{\epsilon(\tau_0)}{\epsilon(\tau_s)}} |\zeta_{k_1}(\tau_0)|^2 |\zeta_{k_2}(\tau_0)|^2 . \label{bsmooth}
\end{align}
Hence, we successfully reproduce bispectrum in \cite{Cai:2018dkf}\footnote{In their notation, $\delta$ and $\epsilon(\tau_s)$ are denoted by $-2 \eta_V$ and $\pi_e^2/2$, respectively.}. We can read that the bispectrum is suppressed by $\delta$, a SR parameter. This suppression comes from the lower bounds of the time integral of total derivative in \eqref{smoothint} which makes the bispectrum proportional to $\eta'(\tau_s)$.

Compared to \cite{Cai:2018dkf}, the explicit form of the mode function \eqref{modeusrsr} is directly substituted for the integral of time \eqref{tintsmooth}. Then, they implemented conditions $k_1 \rightarrow 0$ and $k_2 \ll k_s$, and performed the time integral to obtain \eqref{bsmooth}. In this work, we have obtained a general form of bispectrum without any condition on the wavenumbers given by \eqref{bsmoothfull}. Thus, realizing that the integrand of \eqref{tintsmooth} is a total time derivative, we can get a more general and precise result than \cite{Cai:2018dkf}.

\subsubsection{Slow-roll to ultraslow-roll to slow-roll transitions}
\begin{figure}[tbp]
\centering 
\includegraphics[width=0.75\textwidth]{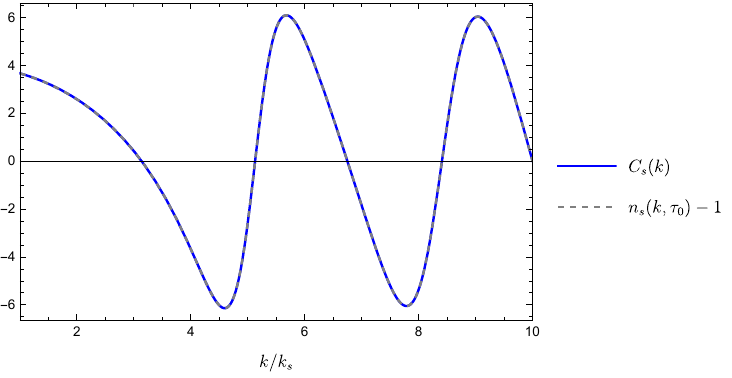}
\caption{\label{fig8} Plot of $C_s(k)$ and $n_s(k,\tau_0) - 1$ for the smooth transition case. We choose $k_e/k_s = 10$ just for illustrative purposes.}
\end{figure}

In this part, we turn back to the previous discussion with two transitions of the second SR parameter: SR to USR and USR back to SR transitions. Contribution of smooth transition from the USR to the final SR period is suppressed by $\delta$ due to the $\eta'(\tau_s)$ prefactor in \eqref{smoothbe}. Therefore, the leading-order bispectrum $\mathcal{O}(\delta^0)$ comes from the first sharp transition
\begin{equation}
\langle\!\langle \zeta_{\bk_1}(\tau_0) \zeta_{\bk_2}(\tau_0) \zeta_{\bk_3}(\tau_0) \rangle\!\rangle = \langle\!\langle \zeta_{\bk_1}(\tau_0) \zeta_{\bk_2}(\tau_0) \zeta_{\bk_3}(\tau_0) \rangle\!\rangle^\mathrm{(s)}.
\end{equation}
In the squeezed limit, $k_1 \rightarrow 0$, the bispectrum becomes
\begin{gather}
\langle \! \langle \zeta_{\bk_1}(\tau_0) \zeta_{\bk_2}(\tau_0) \zeta_{-\bk_2}(\tau_0) \rangle \!\rangle = 4 \Delta\eta \mpl^2 \epsilon(\tau_s) a^2(\tau_s) \mathrm{Im} \left[ \frac{\zeta_{k_2}^2(\tau_0)}{\abs{\zeta_{k_2}(\tau_0)}^2} \zeta_{k_2}^*(\tau_s) \zeta_{k_2}'^*(\tau_s) \right]  \nonumber\\ 
\times \abs{\zeta_{k_1}(\tau_0)}^2 \abs{\zeta_{k_2}(\tau_0)}^2 ,
\end{gather}
where right hand side of the first line is defined as $-C_s(k_2)$. The plot of $C_s(k)$ is shown in Fig. \ref{fig8}. It also shows that Maldacena's theorem is satisfied, namely $C_s(k) = n_s(k, \tau_0) - 1$.

\section{One-loop correction \label{1loop}}
In this section, we calculate one-loop corrections to the large-scale power spectrum generated by cubic and quartic-induced Hamiltonian. We compare the effect of sharp and smooth transition of the second SR parameter to the one-loop corrections. Because we are interested in one-loop correction to the large-scale power spectrum, the operator $\mathcal{O}(\tau_0)$ is taken as $\zeta_\bp(\tau_0) \zeta_{-\bp}(\tau_0)$.

\subsection{Cubic Hamiltonian}
One-loop correction generated by cubic Hamiltonian \eqref{hint} is derived by the second-order expansion of in-in perturbation theory. In Appendix~\ref{app:oneloop}, we show a detailed derivation of the one-loop correction. After some algebra, we obtain
\begin{gather}
\langle\!\langle \zeta_{\bp}(\tau_0) \zeta_{-\bp}(\tau_0) \rangle\!\rangle^{(3)} = \frac{1}{2} \mpl^4 \int_{-\infty}^{\tau_0} \md \tau_1 \epsilon(\tau_1) \eta'(\tau_1) a^2(\tau_1) \int_{-\infty}^{\tau_1} \md \tau_2 \epsilon(\tau_2) \eta'(\tau_2) a^2(\tau_2) \nonumber\\
\times 4 \mathrm{Re} \int \frac{\md^3 k}{(2\pi)^3} \left\lbrace \left( \zeta_p(\tau_0) \zeta_p^*(\tau_0)  \left[ \zeta_p(\tau_1) \zeta_k(\tau_1) \zeta_q(\tau_1) \right]' - \zeta_p(\tau_0) \zeta_p(\tau_0)  \left[ \zeta_p^*(\tau_1) \zeta_k(\tau_1) \zeta_q(\tau_1)  \right]'  \right) \right. \nonumber\\
\times \left. \left[ \zeta_p^*(\tau_2) \zeta_k^*(\tau_2) \zeta_q^*(\tau_2)\right]' \right\rbrace .
\end{gather}
In the next subsection, we will calculate a one-loop correction for a sharp and smooth transition of the second SR parameter.

\subsubsection{Sharp transition}
As in the previous section, for a sharp transition, $\eta'(\tau)$ can be modeled as
\begin{equation}
\eta'(\tau) = \Delta\eta \left[ -\delta(\tau-\tau_s) + \delta(\tau-\tau_e) \right].
\end{equation}
As shown in \cite{Kristiano:2023scm}, contributions from the Delta-dirac function at $\tau = \tau_s$ are subdominant compared to $\tau = \tau_e$. For $p \ll k$, the one-loop correction becomes \cite{Kristiano:2022maq}
\begin{equation}
\langle \! \langle \zeta_{\bp}(\tau_0) \zeta_{-\bp}(\tau_0) \rangle \! \rangle^{(3)} = \frac{1}{4} \mpl^4 (\Delta\eta)^2 \abs{\zeta_p(\tau_0)}^2 \int \frac{\md^3 k}{(2\pi)^3} 16 \left[\epsilon^2  a^4\abs{\zeta_k}^2 \mathrm{Im}(\zeta_p \zeta_p'^*) \mathrm{Im}(\zeta_k \zeta_k'^*) \right]_{\tau = \tau_e}.
\end{equation}
Substituting the normalization condition \eqref{normalization} yields
\begin{equation}
\langle \! \langle \zeta_{\bp}(\tau_0) \zeta_{-\bp}(\tau_0) \rangle \! \rangle^{(3)} = \frac{1}{4} (\Delta\eta)^2 \abs{\zeta_{p}(\tau_0)}^2 \int \frac{\md^3 k}{(2 \pi)^3} \abs{\zeta_{k}(\tau_e)}^2 . \label{loopsharp}
\end{equation}
Therefore, requiring the one-loop correction to be much smaller than the tree-level contribution leads to a constraint
\begin{equation}
\Delta_{s(\mathrm{PBH})}^2 \ll \mathcal{O}(1) \frac{1}{(\Delta\eta)^2} . \label{h3bound}
\end{equation}
More detailed derivation can be found in \cite{Kristiano:2022maq}.

\subsubsection{Smooth transition}\label{sec:smooth}
Using the fact that $\epsilon(\tau) a^2(\tau) \eta'(\tau)$ is a constant during the transition and performing integral with respect to $\tau_2$, the one-loop correction becomes
\begin{gather}
\langle\!\langle \zeta_{\bp}(\tau_0) \zeta_{-\bp}(\tau_0) \rangle\!\rangle^{(3)} = \frac{1}{2} \mpl^4 (\epsilon(\tau_s) \eta'(\tau_s) a^2(\tau_s))^2 \int_{\tau_s}^{\tau_0} \md \tau_1  \int \frac{\md^3 k}{(2\pi)^3} 4 \mathrm{Re} \left\lbrace  \zeta_p^*(\tau_1) \zeta_k^*(\tau_1) \zeta_q^*(\tau_1) \right. \nonumber\\
\times \left.  \left( \zeta_p(\tau_0) \zeta_p^*(\tau_0)  \left[ \zeta_p(\tau_1) \zeta_k(\tau_1) \zeta_q(\tau_1) \right]' - \zeta_p(\tau_0) \zeta_p(\tau_0)  \left[ \zeta_p^*(\tau_1) \zeta_k(\tau_1) \zeta_q(\tau_1)  \right]'  \right) \right\rbrace ,
\end{gather}
After some algebra, the one-loop correction is simplified to
\begin{gather}
\langle\!\langle \zeta_{\bp}(\tau_0) \zeta_{-\bp}(\tau_0) \rangle\!\rangle^{(3)} = \frac{1}{2} \mpl^4 (\epsilon(\tau_s) \eta'(\tau_s) a^2(\tau_s))^2  \\
\times \int_{\tau_s}^{\tau_0} \md \tau_1  \int \frac{\md^3 k}{(2\pi)^3} 4 |\zeta_p(\tau_0)|^2 |\zeta_k(\tau_1)|^2 4 \mathrm{Im}\left[ \zeta_q'(\tau_1) \zeta_q^*(\tau_1) \right] \mathrm{Im}\left[ \zeta_p(\tau_0) \zeta_p^*(\tau_1) \right] . \nonumber
\end{gather}

We want to evaluate the integrand. $\mathrm{Im}\left[ \zeta_q'(\tau_1) \zeta_q^*(\tau_1) \right]$ is simply the normalization condition \eqref{normalization}. For $\abs{p \tau_0} , \abs{p \tau_1} \ll 1$, with the following approximation
\begin{equation}
\frac{\md}{\md \tau_1} \mathrm{Im}\left[ \zeta_p(\tau_0) \zeta_p^*(\tau_1) \right] = \mathrm{Im}\left[ \zeta_p(\tau_0) \zeta_p'^*(\tau_1) \right] \simeq  \mathrm{Im}\left[ \zeta_p(\tau_1) \zeta_p'^*(\tau_1) \right],
\end{equation}
we can obtain
\begin{equation}
\mathrm{Im}\left[ \zeta_p(\tau_0) \zeta_p^*(\tau_1) \right] \simeq \int \frac{\md\tau_1 \eta'(\tau_1)}{4 \mpl^2 \epsilon(\tau_1) a^2(\tau_1) \eta'(\tau_1)} = \frac{\eta(\tau_1)}{4 \mpl^2 \epsilon(\tau_s) a^2(\tau_s) \eta'(\tau_s)} .
\end{equation}
Hence, the one-loop correction becomes
\begin{equation}
\langle\!\langle \zeta_{\bp}(\tau_0) \zeta_{-\bp}(\tau_0) \rangle\!\rangle^{(3)} = - \frac{1}{2} |\zeta_p(\tau_0)|^2 \int \frac{\md^3 k}{(2\pi)^3} \int_{\tau_s}^{\tau_0} \md \tau_1 \eta'(\tau_1) \eta(\tau_1) |\zeta_k(\tau_1)|^2 , \label{smoothtime}
\end{equation}
which can be approximated as
\begin{equation}
\langle\!\langle \zeta_{\bp}(\tau_0) \zeta_{-\bp}(\tau_0) \rangle\!\rangle^{(3)} \simeq \frac{1}{4} (\Delta\eta)^2 |\zeta_p(\tau_0)|^2  \int \frac{\md^3 k}{(2\pi)^3} |\zeta_k(\tau_m)|^2 . \label{loopsmooth}
\end{equation}

Compared to the sharp transition \eqref{loopsharp}, the one-loop correction induced by the cubic Hamiltonian in smooth transition is almost the same. In other words, smoothing the transition cannot significantly suppress the one-loop correction. This statement is supported by numerical studies \cite{Franciolini:2023lgy, Davies:2023hhn}.\footnote{Evolution of the second SR parameter in potential model of \cite{Davies:2023hhn} after the first transition satisfies Wands duality condition. In other words, the quantity $(aH)^{-2} z''/z$ remains constant throughout the USR and the following phases, with evolution similar to the blue curve in Fig.~\ref{figm}. They find that the one-loop correction generated by the cubic Hamiltonian in the potential model (smooth transition) is almost equal to the instantenous model (sharp transition), which is consistent with our analytical prediction. We thank the authors of \cite{Davies:2023hhn} for private communication.} Although the bispectrum is suppressed by a small parameter $\delta$, the suppression does not appear in the one-loop correction. Contrary to the bispectrum that is evaluated at the boundary of the time integral, the one-loop correction has a contribution from the bulk time. It is indicated by the fact that the time integral in \eqref{smoothtime} cannot be expressed as an integral of a total derivative, which comes from the fact that the second-order in-in perturbation theory contains a time-ordered integral. 

This result also implies that by looking only at the ``coupling constant'' one cannot accurately estimate the one-loop correction precisely. From \eqref{hint}, one might think that small $\eta'$ yields a negligible one-loop correction. However, it is an incorrect expectation because one needs to perform a time integral to obtain the one-loop correction. Although $\eta'$ is small, it can generate a large one-loop correction if the duration is long enough. The one-loop correction can be comparable to a case where $\eta'$ is large but only for a short period. However, we should not regard \eqref{loopsmooth} as the final result, for the reason we will explain in the next subsection.

Another remark is that a small contribution to the bispectrum does not imply a small contribution to the one-loop correction. For the sharp transition case, we can see in Fig.~\ref{fig7} that the contribution from the time integral at $\tau = \tau_e$ to the bispectrum denoted by $C_e(k)$ is much smaller than $C_s(k)$. However, the time integral at $\tau = \tau_e$ contributes to the one-loop correction much more significantly than $\tau = \tau_s$. Similar situation for the smooth transition case. We can see in Fig.~\ref{fig8} that the smooth transition at $\tau > \tau_s$ does not contribute to the leading-order bispectrum because it is suppressed by $\delta$. However, the one-loop correction \eqref{loopsmooth} is not suppressed by this parameter.

\subsection{Quartic-induced Hamiltonian}
Consider a second-order, third-order, and fourth-order Lagrangian with form
\begin{align}
\mathcal{L}_2 & = f_0 \dot{\zeta}^2 + j_2 \nonumber\\
\mathcal{L}_3 & = g_0 \dot{\zeta}^3 + g_1 \dot{\zeta}^2 + g_2 \dot{\zeta} + j_3 \nonumber\\
\mathcal{L}_4 & = h_0 \dot{\zeta}^4 + h_1 \dot{\zeta}^3 + h_2 \dot{\zeta}^2 + h_3 \dot{\zeta} + j_4,
\end{align}
where all coefficients of $\dot{\zeta}$ are a function of $\zeta$ and its spatial derivative. In addition to the cubic Hamiltonian $\mathcal{H}^{(3)} = - \mathcal{L}_3$, the Lagrangian induces the following fourth-order Hamiltonian \cite{Chen:2009bc}
\begin{align}
\mathcal{H}^\mathrm{(4I)} = & \left( \frac{9 g_0^2}{4 f_0 } - h_0 \right) \dot{\zeta}^4 + \left( \frac{3 g_0 g_1}{f_0} - h_1 \right) \dot{\zeta}^3 + \left( \frac{3 g_0 g_2}{2f_0} + \frac{g_1^2}{f_0} - h_2 \right) \dot{\zeta}^2 \nonumber\\
& + \left( \frac{g_1 g_2}{f_0} - h_3 \right) \dot{\zeta} + \frac{g_2^2}{4f_0} - j_4.
\end{align}
Hence, the last term in \eqref{s3bulk} generates
\begin{equation}
H^\mathrm{(4I)}(\tau) = \frac{1}{16} \mpl^2 \int \md^3 x ~a^2(\tau) \epsilon(\tau) \eta'^2(\tau) \zeta^4(\bx,\tau), \label{4ind}
\end{equation}
which is called quartic-induced Hamiltonian.
\subsubsection{Smooth transition}
Combining second-order and third-order action reads
\begin{equation}
S[\zeta] = M_{\mathrm{pl}}^2 \int \mathrm{d}\tau \mathrm{d}^3x ~a^2 \epsilon  \left[ (\zeta')^2 - (\partial_i \zeta)^2 + \frac{1}{2} \eta' \zeta'\zeta^2 \right],
\end{equation}
which has equation of motion in momentum space
\begin{equation}
\zeta_\bp'' + \frac{(a^2 \epsilon)'}{a^2 \epsilon} \zeta_\bp' + p^2 \zeta_\bp = - \frac{(a^2 \epsilon \eta')'}{4a^2 \epsilon} \int \frac{\md^3 k}{(2\pi)^3} \zeta_\bk \zeta_{\bp-\bk} . \label{eom}
\end{equation}
We can easily read that if $\epsilon(\tau) a^2(\tau) \eta'(\tau) = \mathrm{constant}$, the right-hand side of the equation of motion becomes zero. This means that the equation of motion is the same as the free-theory equation of motion without higher-order interactions. Thus, the bulk time contribution to the correlation function in \eqref{smoothtime} does not make sense if it is regarded as the final result. There is a missing contribution from the quartic-induced Hamiltonian \eqref{4ind} that needs to be added.

In Appendix~\ref{app:total}, we show that a third-order action with total time derivative of $\zeta^3$ yields a vanishing correlation function of $\zeta$ at the second-order expansion of the in-in perturbation theory. Thus, we can see cancelation between one-loop correction generated by the cubic and quartic-induced Hamiltonian;
\begin{equation}
\langle\!\langle \zeta_{\bp}(\tau_0) \zeta_{-\bp}(\tau_0) \rangle\!\rangle^{(3)} + \langle\!\langle \zeta_{\bp}(\tau_0) \zeta_{-\bp}(\tau_0) \rangle\!\rangle^\mathrm{(4I)} = 0,
\end{equation}
which is consistent with equation of motion \eqref{eom} viewpoint. This implies that there is no $\mathcal{O}(\delta^0)$ contribution to the one-loop correction. The leading contribution to the one-loop correction is expected to have $\delta$ suppression, which comes from the contribution at the boundary $\tau = \tau_s$. Therefore, for a smooth transition that satisfies the Wands duality condition, the one-loop correction is suppressed by $\delta$. Note that this exact cancellation only happens when transition of the second SR parameter satisfies the Wands duality condition. For other function of $\eta(\tau)$, even if it is smooth or continuously evolves, one should not expect such an exact cancellation.

We can show that the Wands duality condition has a stronger implication, which is a free theory at all orders of perturbation. Consider the relevant third-order action \footnote{Strictly speaking, $\zeta$ in this paragraph should be $\bz$ defined in Sect.~\ref{3pt}.}
\begin{equation}
S_{\mathrm{bulk}}[\zeta] \supset \frac{1}{2} \mpl^2  \int \md t ~\md^3 x ~ a^3 \epsilon \dot{\eta} \dot{\zeta} \zeta^2 = \frac{1}{6} \mpl^2 \int \md t ~\md^3 x \left[ \partial_t \left(  a^3 \epsilon \dot{\eta} \zeta^3 \right) - \partial_t \left( a^3 \epsilon \dot{\eta} \right) \zeta^3 \right].
\end{equation}
As shown in Appendix~\ref{app:total}, the total time derivative interaction does not contribute to the correlation of $\zeta$, so it can be safely neglected. Transforming into a flat-slicing gauge $\zeta = - \delta\phi / (\mpl \sqrt{2 \epsilon})$, the third-order action becomes
\begin{equation}
S_{\mathrm{bulk}}[\delta\phi] = - \frac{1}{6} \mpl^2 \int \md t ~\md^3 x ~\partial_t \left( a^3 \epsilon \dot{\eta} \right) \frac{\delta\phi^3}{( \mpl \sqrt{2\epsilon} )^3} .
\end{equation}
In terms of third derivative of the potential, the third-order action can be expressed as
\begin{equation}
S_{\mathrm{bulk}}[\delta\phi] = - \frac{1}{6} \int \md t ~\md^3 x ~a^3 V_3 \delta\phi^3 ,
\end{equation}
where
\begin{equation}
V_n \equiv \frac{\md^n V}{\md \phi^n} ~;\mathrm{for}~ n \in \mathbb{Z},
\end{equation}
\begin{equation}
V_3 = - \frac{\mpl^2}{( \mpl a\sqrt{2\epsilon} )^3} \partial_t \left( a^3 \epsilon \dot{\eta} \right)  = - \frac{H}{\mpl \sqrt{2\epsilon}} \partial_t \left( \nu^2 - \frac{9}{4} \right).
\end{equation}
Because Wands duality condition is characterized by a constant $\nu$, the third derivative of the potential vanishes. Then, the higher-order derivative of the potential can be written as $V_{n+1} = \dot{V}_n / \dot{\phi}$, so the vanishing third derivative implies vanishing derivative of the potential with order greater than three.

\subsubsection{Sharp transition}
We have shown that the one-loop correction generated by cubic Hamiltonian and quartic-induced Hamiltonian cancel with each other if the transition of the second SR parameter satisfies the Wands duality condition. Then, one might wonder the effect of quartic-induced Hamiltonian to the one-loop correction in the case of sharp transition. The first-order in-in perturbation theory reads
\begin{equation}
\langle \zeta_{\bp}(\tau_0) \zeta_{-\bp}(\tau_0) \rangle^\mathrm{(4I)} = 2 \int_{-\infty}^{\tau_0} \mathrm{d}\tau_1 ~\mathrm{Im} \left\langle \zeta_{\bp}(\tau_0) \zeta_{-\bp}(\tau_0) H^\mathrm{(4I)}(\tau_1) \right\rangle .
\end{equation}
Substituting the quartic-induced Hamiltonian \eqref{4ind} yields
\begin{gather}
\langle \zeta_{\bp}(\tau_0) \zeta_{-\bp}(\tau_0) \rangle^\mathrm{(4I)} = \frac{2}{16} \mpl^2 \int_{-\infty}^{\tau_0} \mathrm{d}\tau_1 a^2(\tau_1) \epsilon(\tau_1) \eta'(\tau_1) \eta'(\tau_1) \\
\times \int \prod_{a=1}^4  \left[ \frac{\md^3 k_a}{(2\pi)^3} \right] \delta(\bk_1 + \bk_2 + \bk_3 + \bk_4) \mathrm{Im} \langle \zeta_{\bp}(\tau_0) \zeta_{-\bp}(\tau_0) \zeta_{\bk_1}(\tau_1) \zeta_{\bk_2}(\tau_1) \zeta_{\bk_3}(\tau_1) \zeta_{\bk_4}(\tau_1)  \rangle . \nonumber
\end{gather}
Performing Wick contraction leads to
\begin{align}
\langle\!\langle \zeta_{\bp}(\tau_0) \zeta_{-\bp}(\tau_0) \rangle\!\rangle^\mathrm{(4I)} =& ~\frac{2}{16} \mpl^2 \int_{-\infty}^{\tau_0} \mathrm{d}\tau_1 ~a^2(\tau_1) \epsilon(\tau_1) \eta'(\tau_1)  \eta'(\tau_1) \\
& \times 24 \int \frac{\md^3k}{(2\pi)^3} |\zeta_p(\tau_0)|^2 \mathrm{Im}[ \zeta_p(\tau_0) \zeta_p^*(\tau_1) ] |\zeta_k(\tau_1)|^2 . \nonumber
\end{align}
Then, we want to evaluate the time integral. For the sharp transition case, $\eta'(\tau)$ is modeled as a Dirac delta function. Evaluating integral of a Dirac-delta function yields
\begin{align}
\langle\!\langle \zeta_{\bp}(\tau_0) \zeta_{-\bp}(\tau_0) \rangle\!\rangle^\mathrm{(4I)} =& ~\frac{2}{16} \mpl^2 a^2(\tau_e) \epsilon(\tau_e) \Delta\eta ~\eta'(\tau_e) \\
& \times 24 \int \frac{\md^3k}{(2\pi)^3} |\zeta_p(\tau_0)|^2 \mathrm{Im}[ \zeta_p(\tau_0) \zeta_p^*(\tau_e) ] |\zeta_k(\tau_e)|^2 . \nonumber
\end{align}
We find that the one-loop correction is proportional to $\eta'(\tau_e) = \Delta\eta ~\delta(0)$ that is infinity. However, this divergence appears to be due to an idealized model. In reality, from a generic inflationary potential, the transition of the second SR parameter should have a finite width, $\Delta\tau$. We can regulate $\delta(0)$ as $1/\Delta\tau$, thus
\begin{equation}
\eta'(\tau_e) = \lim_{\Delta\tau \rightarrow 0} \frac{\Delta\eta}{\Delta\tau} .
\end{equation}
Note that this regularization is performed only for the time integral and has nothing to do with the momentum integral.

From the mode function \eqref{zetasr2}, we obtain
\begin{equation}
\mathrm{Im}[ \zeta_p(\tau_0) \zeta_p^*(\tau_e) ] = -  \frac{H^2}{12 M_{\mathrm{pl}}^2 \epsilon(\tau_s) p^3}  \left( \frac{p}{k_s} \right)^3 \left( \frac{k_e}{k_s} \right)^3. 
\end{equation}
Substituting it to the one-loop correction yields
\begin{equation}
\langle\!\langle \zeta_{\bp}(\tau_0) \zeta_{-\bp}(\tau_0) \rangle\!\rangle^\mathrm{(4I)} = \frac{1}{4} (\Delta\eta)^2 \frac{\tau_e}{\Delta\tau} |\zeta_p(\tau_0)|^2 \int \frac{\md^3k}{(2\pi)^3} |\zeta_k(\tau_e)|^2 , \label{4indloop}
\end{equation}
Then, performing momentum integral from $k_s$ to $k_e$ yields one-loop correction to the power spectrum
\begin{equation}
\Delta_{s(1)}^2(p, \tau_0) = \frac{1}{4} (\Delta\eta)^2 \frac{\tau_e}{\Delta\tau} \Delta_{s(0)}^2(p, \tau_0) \Delta_{s(\mathrm{PBH})}^2 .
\end{equation}
Hence, requiring one-loop correction to be much smaller than tree-level contribution leads to a constraint
\begin{equation}
 \Delta_{s(\mathrm{PBH})}^2 \ll \mathcal{O}(1) \frac{1}{(\Delta\eta)^2} \frac{\Delta\tau}{\tau_e} , \label{h4bound}
\end{equation}
which is a much stronger condition than \eqref{h3bound} because $\Delta\tau/\tau_e \rightarrow 0$. In Appendix~\ref{app:source}, we derive this one-loop correction from quartic-induced Hamiltonian by another method called source method, which directly solves the equation of motion with self-interaction term.

\section{Conclusion \label{con}}

In this paper, we have elaborated the differences in tree-level power spectrum, bispectrum, and one-loop correction to the power spectrum between inflation models with sharp and smooth transitions of the second SR parameter from USR to SR period. When we published \cite{Kristiano:2022maq}, many people wondered if the large one-loop correction we found was due to our assumption of discontinuous evolution of the second SR parameter corresponding to the sharp transition. In Sect.~\ref{sec:smooth}, however, we have shown that even when the transition is smooth, we suffer from one-loop corrections of the same order of magnitude for the same one-loop diagrams as we studied in our previous paper in the case of the sharp transition \cite{Kristiano:2022maq}. However, this is not the end of the story, since we also need to consider the effects generated by the quartic-induced Hamiltonian.

To describe a smooth transition analytically, as we did in the present manuscript, a model satisfying (\ref{etadif}) or $(\epsilon a^2 \eta' )' =0$ is often adopted, which is equivalent to the case the Wands duality condition is satisfied. In fact, this model has a free parameter $\delta$, which control sharpness of the transition by the value of $\eta'$ at its peak. Thus, rather than calling such a transition as smooth transition, it is better to specify it as transition satisfying Wands duality condition.

We have found that the Wands duality condition makes the important cubic self-interaction to be a total time derivative term, which leads to a vanishing leading-order bispectrum. As a result, the bispectrum is suppressed by a small parameter $\delta$, which is the second SR parameter at the end of inflation. This result is consistent with a previous reference \cite{Cai:2018dkf}. We would like to emphasize that a suppressed bispectrum is not a result of {\it any} smooth evolution of the second SR parameter but rather due to the particular choice of transition satisfying the Wands duality. So, it should not be regarded as a generic example of smooth transition.

As we argued in Sect.~4, there are two contributions to the one-loop correction: a cubic Hamiltonian and a quartic-induced Hamiltonian. For the one-loop correction induced by the cubic Hamiltonian, we have shown that the sharp and smooth transitions yield almost equal results. This is consistent with the numerical result in previous references \cite{Iacconi:2023ggt}. However, we have demonstrated that the contribution from the quartic-induced Hamiltonian exactly cancels the one induced by the cubic Hamiltonian for a transition that satisfies the Wands duality condition. In case of a sharp transition where the time derivative of the second SR parameter is modeled by a Dirac-delta function of time, we have found a divergent contribution to the one-loop correction induced by quartic-induced Hamiltonian. However, the divergence appears to be due to the idealized model of the transition. If the Dirac-delta function is regularized by introducing a finite width of transition duration, we have found that the one-loop correction is inversely proportional to the transition duration. As a caveat, there should be additional contributions to the one-loop correction by quartic Hamiltonian from fourth-order action of the curvature perturbation. The term in the one-loop correction that is inversely proportional to the duration of the transition is not found in \cite{Firouzjahi:2023aum} based on the effective field theory approach \cite{Akhshik:2015nfa}, so we expect there should be a cancelation with the quartic Hamiltonian.\footnote{We will show this cancellation in the future.} Calculating such contributions is beyond the scope of this paper, as we focus on the effect of sharp and smooth transitions of the second SR parameter to the bispectrum and one-loop correction to the power spectrum generated by the relevant cubic self-interaction of the curvature perturbation.

\begin{figure}[tbp]
\centering 
\includegraphics[width=0.5\textwidth]{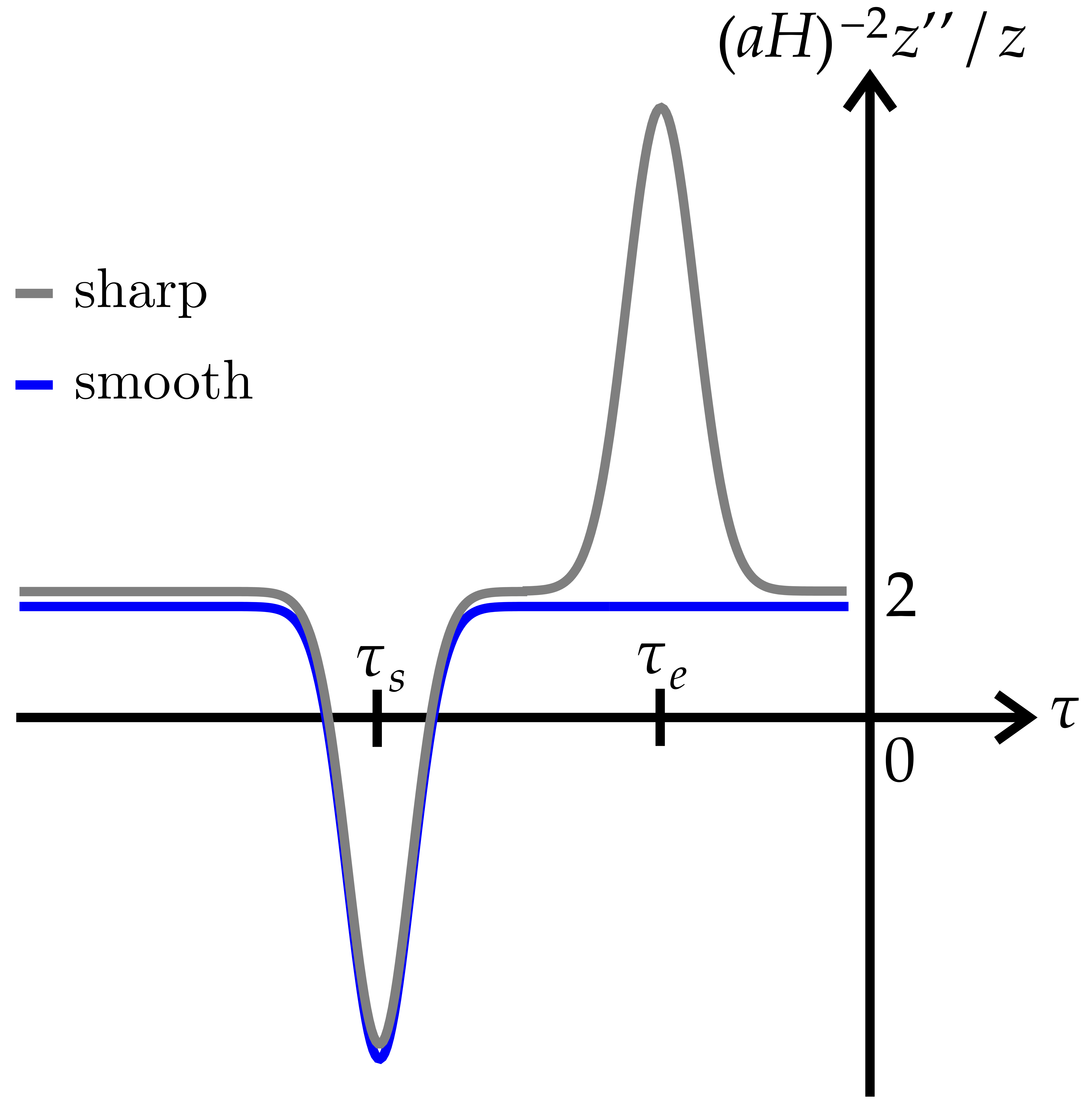}
\caption{\label{figm} The sketch of the evolution of the effective mass $(aH)^{-2} z''/z$ corresponds to the evolution of the second SR parameter in Fig.~\ref{fig1}. The peak can be modeled as a Dirac delta function of time. }
\end{figure}

In conclusion, we have shown that the Wands duality condition on the transition of the second SR parameter yields a vanishing one-loop correction to the power spectrum. Other smooth evolutions of the second SR parameter do not lead to a vanishing one-loop correction. Our result implies that extreme fine tuning of the inflationary potential is necessary to make it satisfy the Wands duality condition so that higher-order correlation functions become negligible. In the sharp-transition case, the one-loop correction is extremely large, and thus it casts doubt on the validity of perturbation theory. Rather than focusing  on the evolution of the second SR parameter, a simpler way to deduce whether the model at hand has large one-loop correction is to read the evolution of $(aH)^{-2} z''/z$, as shown in Fig.~\ref{figm}. The Wands duality condition corresponds to a constant value of $(aH)^{-2} z''/z$. If it stays constant after the first transition, as in the blue curve, the model is free from a large one-loop correction. However, if it has a strong time dependence after the first transition, like the gray curve, the model suffers from large one-loop correction.\footnote{We thank David Wands for pointing out this method.}

Finally, we would like to ask an open question: is there any deep insight behind vanishing one-loop correction due to the Wands duality condition? In particular, it is curious to read that imposing the Wands duality condition on the linear equation of motion makes the relevant cubic self-interaction a topological term. Moreover, the Wands duality condition implies that all leading higher-order self-interactions of the curvature perturbation vanish, making the linear perturbation theory an almost perfect approximation. Deviation from the Wands duality condition parameterizes how large the nonlinear correction to the correlation function is. Our result might shed light on the bootstrap approach for cosmological correlators in an inflation model with violation of the SR approximation. In standard SR inflation, cosmological correlators can be bootstrapped by considering slightly broken dS isometries due to the SR parameters \cite{Arkani-Hamed:2018kmz}. In case dS boost isometry is strongly broken that yields large non-Gaussianity, the guiding principles to boostrap cosmological correlators are explained in \cite{Pajer:2020wxk}. Thus, we speculate that deviation from the Wands duality condition might be a guiding principle to bootstrap cosmological correlators in an inflation model with violation of SR approximation. We hope to discuss this idea more in the future.

\section*{Notes Added}
While we were writing this paper, Refs. \cite{Braglia:2024zsl, Ballesteros:2024zdp} appeared on arXiv with consistent overlapping results for this paper. It is shown in Ref. \cite{Braglia:2024zsl} that $\cb(\zeta) \propto \zeta^3$ yields a vanishing correlation function of $\zeta$ in second-order perturbation theory because the contributions from cubic and quartic-induced Hamiltonian cancel each other out. A novelty of this paper is that we have shown that the Wands duality condition leads to such a result. In the sharp transition case, a divergent contribution to the one-loop correction from the quartic-induced Hamiltonian is found by Ref. \cite{Ballesteros:2024zdp}. While we agree that it is the dominant contribution generated by the relevant third-order action in sharp transition case, another novelty of this paper is that we have shown an example in which the cubic and quartic-induced Hamiltonian generate comparable contributions to the one-loop correction and cancel with each other.

\acknowledgments
J.~K. acknowledges the support from JSPS KAKENHI Grants No.22KJ1006 and No.22J20289. J.~K. is also supported by the Global Science Graduate Course (GSGC) program of The University of Tokyo. J.~Y. is supported by JSPS KAKENHI Grant No.~20H05639. This research was supported by the Munich Institute for Astro-, Particle and BioPhysics (MIAPbP) which is funded by the Deutsche Forschungsgemeinschaft (DFG, German Research Foundation) under Germany's Excellence Strategy – EXC-2094 – 390783311. We thank David Wands for an interesting discussion in the MIAPbP workshop ``Quantum Aspects of Inflationary Cosmology''.

\appendix
\section{Derivation of the one-loop correction \label{app:oneloop}}
The cubic Hamiltonian \eqref{hint} contributes to a one-loop correction to the power spectrum in second-order perturbation theory. The second-order expansion of in-in perturbation theory reads
\begin{gather}
\langle \mathcal{O}(\tau) \rangle^{(3)} = \langle \mathcal{O}(\tau) \rangle_{(1,1)} + 2 \mathrm{Re} \langle \mathcal{O}(\tau) \rangle_{(0,2)}, \label{o3} \\
\langle \mathcal{O}(\tau) \rangle_{(1,1)} = \int_{-\infty}^{\tau} \mathrm{d}\tau_1 \int_{-\infty}^{\tau} \mathrm{d}\tau_2 \left\langle H^{(3)}(\tau_1) \mathcal{\hat{O}} (\tau) H^{(3)}(\tau_2) \right\rangle, \nonumber\\
\langle \mathcal{O}(\tau) \rangle_{(0,2)} = - \int_{-\infty}^{\tau} \mathrm{d}\tau_1 \int_{-\infty}^{\tau_1} \mathrm{d}\tau_2 \left\langle \mathcal{\hat{O}} (\tau) H^{(3)}(\tau_1) H^{(3)}(\tau_2) \right\rangle. \nonumber
\end{gather}
Performing Wick contraction, the $(1,1)$ and $(0,2)$ terms become
\begin{gather}
\langle\!\langle \zeta_{\bp}(\tau_0) \zeta_{-\bp}(\tau_0) \rangle\!\rangle_{(1,1)} =  \frac{1}{4} \mpl^4 \zeta_p(\tau_0) \zeta_p^*(\tau_0) \int_{-\infty}^{\tau_0} \md \tau_1 \epsilon(\tau_1) \eta'(\tau_1) a^2(\tau_1) \int_{-\infty}^{\tau_0} \md \tau_2 \epsilon(\tau_2) \eta'(\tau_2) a^2(\tau_2) \nonumber\\
\times \int \frac{\md^3 k}{(2\pi)^3} \left[ 4 \zeta_p'(\tau_1) \zeta_k(\tau_1) \zeta_q(\tau_1 ) \zeta_p'^*(\tau_2) \zeta_k^*(\tau_2) \zeta_q^*(\tau_2 ) + 8 \zeta_p(\tau_1) \zeta_k'(\tau_1) \zeta_q(\tau_1 ) \zeta_p^*(\tau_2) \zeta_k^*(\tau_2) \zeta_q'^*(\tau_2 ) \right. \nonumber\\
\left. + 8 \zeta_p(\tau_1) \zeta_k'(\tau_1) \zeta_q(\tau_1 ) \zeta_p^*(\tau_2) \zeta_k'^*(\tau_2) \zeta_q^*(\tau_2 ) + 8 \zeta_p'(\tau_1) \zeta_k(\tau_1) \zeta_q(\tau_1 ) \zeta_p^*(\tau_2) \zeta_k^*(\tau_2) \zeta_q'^*(\tau_2 ) \right. \nonumber\\
\left. + 8 \zeta_p(\tau_1) \zeta_k'(\tau_1) \zeta_q(\tau_1 ) \zeta_p'^*(\tau_2) \zeta_k^*(\tau_2) \zeta_q^*(\tau_2 ) \right] ,
\end{gather}
\begin{gather}
\langle\!\langle \zeta_{\bp}(\tau_0) \zeta_{-\bp}(\tau_0) \rangle\!\rangle_{(0,2)} =  -\frac{1}{4} \mpl^4 \zeta_p(\tau_0) \zeta_p(\tau_0) \int_{-\infty}^{\tau_0} \md \tau_1 \epsilon(\tau_1) \eta'(\tau_1) a^2(\tau_1) \int_{-\infty}^{\tau_1} \md \tau_2 \epsilon(\tau_2) \eta'(\tau_2) a^2(\tau_2) \nonumber\\
\times \int \frac{\md^3 k}{(2\pi)^3} \left[ 4 \zeta_p'^*(\tau_1) \zeta_k(\tau_1) \zeta_q(\tau_1 ) \zeta_p'^*(\tau_2) \zeta_k^*(\tau_2) \zeta_q^*(\tau_2 ) + 8 \zeta_p^*(\tau_1) \zeta_k'(\tau_1) \zeta_q(\tau_1 ) \zeta_p^*(\tau_2) \zeta_k^*(\tau_2) \zeta_q'^*(\tau_2 ) \right. \nonumber\\
\left. + 8 \zeta_p^*(\tau_1) \zeta_k'(\tau_1) \zeta_q(\tau_1 ) \zeta_p^*(\tau_2) \zeta_k'^*(\tau_2) \zeta_q^*(\tau_2 ) + 8 \zeta_p'^*(\tau_1) \zeta_k(\tau_1) \zeta_q(\tau_1 ) \zeta_p^*(\tau_2) \zeta_k^*(\tau_2) \zeta_q'^*(\tau_2 ) \right. \nonumber\\
\left. + 8 \zeta_p^*(\tau_1) \zeta_k'(\tau_1) \zeta_q(\tau_1 ) \zeta_p'^*(\tau_2) \zeta_k^*(\tau_2) \zeta_q^*(\tau_2 ) \right] .
\end{gather}
After some algebra, they can be written in terms of total time derivative that read
\begin{align}
\langle\!\langle \zeta_{\bp}(\tau_0) \zeta_{-\bp}(\tau_0) \rangle\!\rangle_{(1,1)} = & ~\frac{1}{4} \mpl^4 \zeta_p(\tau_0) \zeta_p^*(\tau_0) \int_{-\infty}^{\tau_0} \md \tau_1 \epsilon(\tau_1) \eta'(\tau_1) a^2(\tau_1) \int_{-\infty}^{\tau_0} \md \tau_2 \epsilon(\tau_2) \eta'(\tau_2) a^2(\tau_2) \nonumber\\
& \times \int \frac{\md^3 k}{(2\pi)^3} 4 \left[ \zeta_p(\tau_1) \zeta_k(\tau_1) \zeta_q(\tau_1) \right]' \left[ \zeta_p^*(\tau_2) \zeta_k^*(\tau_2) \zeta_q^*(\tau_2)\right]',
\end{align}
\begin{align}
\langle\!\langle \zeta_{\bp}(\tau_0) \zeta_{-\bp}(\tau_0) \rangle\!\rangle_{(0,2)} = & ~\frac{1}{4} \mpl^4 \zeta_p(\tau_0) \zeta_p(\tau_0) \int_{-\infty}^{\tau_0} \md \tau_1 \epsilon(\tau_1) \eta'(\tau_1) a^2(\tau_1) \int_{-\infty}^{\tau_1} \md \tau_2 \epsilon(\tau_2) \eta'(\tau_2) a^2(\tau_2) \nonumber\\
& \times \int \frac{\md^3 k}{(2\pi)^3} 4 \left[ \zeta_p^*(\tau_1) \zeta_k(\tau_1) \zeta_q(\tau_1) \right]' \left[ \zeta_p^*(\tau_2) \zeta_k^*(\tau_2) \zeta_q^*(\tau_2)\right]'. \label{inin02}
\end{align}
The $(1,1)$ term can be further manipulated to be time-ordered integral
\begin{gather}
\langle\!\langle \zeta_{\bp}(\tau_0) \zeta_{-\bp}(\tau_0) \rangle\!\rangle_{(1,1)} = \frac{1}{2} \mpl^4 \zeta_p(\tau_0) \zeta_p^*(\tau_0) \int_{-\infty}^{\tau_0} \md \tau_1 \epsilon(\tau_1) \eta'(\tau_1) a^2(\tau_1) \int_{-\infty}^{\tau_1} \md \tau_2 \epsilon(\tau_2) \eta'(\tau_2) a^2(\tau_2) \nonumber\\
\times \int \frac{\md^3 k}{(2\pi)^3} 4 \mathrm{Re} \left\lbrace \left[ \zeta_p(\tau_1) \zeta_k(\tau_1) \zeta_q(\tau_1) \right]' \left[ \zeta_p^*(\tau_2) \zeta_k^*(\tau_2) \zeta_q^*(\tau_2)\right]' \right\rbrace . \label{inin11}
\end{gather}
Until this point, we have not performed any approximation. The total one-loop correction is given by
\begin{equation}
\langle \! \langle \zeta_{\bp}(\tau_0) \zeta_{-\bp}(\tau_0) \rangle \! \rangle^{(3)} = \langle\!\langle \zeta_{\bp}(\tau_0) \zeta_{-\bp}(\tau_0) \rangle\!\rangle_{(1,1)} + 2 \mathrm{Re} \langle\!\langle \zeta_{\bp}(\tau_0) \zeta_{-\bp}(\tau_0) \rangle\!\rangle_{(0,2)} . \label{looptot}
\end{equation}
From \eqref{inin02}, \eqref{inin11}, and \eqref{looptot}, the one-loop correction can be further simplified as
\begin{gather}
\langle\!\langle \zeta_{\bp}(\tau_0) \zeta_{-\bp}(\tau_0) \rangle\!\rangle^{(3)} = \frac{1}{2} \mpl^4 \int_{-\infty}^{\tau_0} \md \tau_1 \epsilon(\tau_1) \eta'(\tau_1) a^2(\tau_1) \int_{-\infty}^{\tau_1} \md \tau_2 \epsilon(\tau_2) \eta'(\tau_2) a^2(\tau_2) \nonumber\\
\times 4 \mathrm{Re} \int \frac{\md^3 k}{(2\pi)^3} \left\lbrace \left( \zeta_p(\tau_0) \zeta_p^*(\tau_0)  \left[ \zeta_p(\tau_1) \zeta_k(\tau_1) \zeta_q(\tau_1) \right]' - \zeta_p(\tau_0) \zeta_p(\tau_0)  \left[ \zeta_p^*(\tau_1) \zeta_k(\tau_1) \zeta_q(\tau_1)  \right]'  \right) \right. \nonumber\\
\times \left. \left[ \zeta_p^*(\tau_2) \zeta_k^*(\tau_2) \zeta_q^*(\tau_2)\right]' \right\rbrace .
\end{gather}

\section{Total time derivative interaction \label{app:total}}
Suppose we have a third-order action with a total time derivative Lagrangian
\begin{equation}
S_\cb^{(3)}[\zeta] = \int \md\tau ~\md^3x  ~\cb'(\zeta(\bx,\tau)),
\end{equation}
with the corresponding Hamiltonian
\begin{equation}
H_\cb^{(3)} (\tau) = - \int \md^3x  ~\cb'(\zeta(\bx,\tau)).
\end{equation}
We would like to find its effect to correlation functions of $\zeta$. Consider an operator $\mathcal{\hat{O}}$ as a function of $\zeta$. The $(1,1)$ term can easily be simplified to
\begin{align}
\langle \mathcal{O}(\tau_0) \rangle_{(1,1)} &= \int_{-\infty}^{\tau_0} \mathrm{d}\tau_1 \int_{-\infty}^{\tau_0} \mathrm{d}\tau_2 \langle H_\cb^{(3)}(\tau_1) \mathcal{\hat{O}} (\tau_0) H_\cb^{(3)}(\tau_2) \rangle \nonumber\\
&= \int \md^3x \md^3y  \int_{-\infty}^{\tau_0} \mathrm{d}\tau_1 \int_{-\infty}^{\tau_0} \mathrm{d}\tau_2 \left\langle  \cb'( \zeta(\bx, \tau_1)) \mathcal{\hat{O}} (\tau_0)  \cb'( \zeta(\by, \tau_2)) \right\rangle \nonumber\\
&= \int \md^3x \md^3y \left\langle \cb( \zeta(\bx, \tau_0))  \mathcal{\hat{O}} (\tau_0) \cb( \zeta(\by, \tau_0)) \right\rangle , \label{o311}
\end{align}
while the $(0,2)$ term becomes
\begin{align}
 \langle \mathcal{O}(\tau_0) \rangle_{(0,2)} &= - \int_{-\infty}^{\tau_0} \mathrm{d}\tau_1 \int_{-\infty}^{\tau_1} \mathrm{d}\tau_2 ~ \langle \mathcal{\hat{O}} (\tau_0) H_\cb^{(3)}(\tau_1) H_\cb^{(3)}(\tau_2) \rangle \\
&= - \int \md^3x \md^3y \int_{-\infty}^{\tau_0} \mathrm{d}\tau_1 \int_{-\infty}^{\tau_1} \mathrm{d}\tau_2 ~\left\langle \mathcal{\hat{O}} (\tau_0)  \cb'( \zeta(\bx, \tau_1))  \cb'( \zeta(\by, \tau_2)) \right\rangle \nonumber\\
&= - \int \md^3x \md^3y \int_{-\infty}^{\tau_0} \mathrm{d}\tau_1  ~ \left\langle \mathcal{\hat{O}} (\tau_0)  \cb'( \zeta(\bx, \tau_1))   \cb( \zeta(\by, \tau_1)) \right\rangle . \nonumber
\end{align}
Due to time-ordered integral in the $(0,2)$ term, there is a remaining time integral. Performing integration by part, the $(0,2)$ term becomes
\begin{align}
\langle \mathcal{O}(\tau_0) \rangle_{(0,2)} = & -\frac{1}{2} \int \md^3x \md^3y \left\langle \mathcal{\hat{O}} (\tau_0)  \cb( \zeta(\bx, \tau_0)) \cb( \zeta(\by, \tau_0)) \right\rangle \label{o302} \\
& -\frac{1}{2} \int \md^3x \md^3y \int_{-\infty}^{\tau_0} \mathrm{d}\tau_1 \left\langle \mathcal{\hat{O}} (\tau_0) \left[ \cb'( \zeta(\bx, \tau_1)) ,  \cb( \zeta(\by, \tau_1)) \right] \right\rangle \nonumber.
\end{align}
Then, the total one-loop correction \eqref{o3} reads
\begin{align}
\langle \mathcal{O}(\tau_0) \rangle^{(3)} = - \int \md^3x \md^3y \int_{-\infty}^{\tau_0} \mathrm{d}\tau_1 ~\mathrm{Re} \left\langle \mathcal{\hat{O}} (\tau_0) \left[ \cb'( \zeta(\bx, \tau_1)) ,  \cb( \zeta(\by, \tau_1)) \right] \right\rangle, \label{o3bulk}
\end{align}
where the $(1,1)$ term \eqref{o311} is cancelled by two times the first line in the $(0,2)$ term \eqref{o302}. We can read that the remaining contribution to the one-loop correction is a bulk time integral, which cannot be reduced to an integral of total derivative. After substituting $\mathcal{\hat{O}} = \zeta^2$ and transforming to momentum space, the one-loop correction \eqref{o3bulk} corresponds to \eqref{smoothtime}.

Substituting the explicit form of $\cb( \zeta(\bx, \tau)) = c(\tau) \zeta^3(\bx, \tau) $ leads to
\begin{align}
\int \md^3x \md^3y \left[ \cb'( \zeta(\bx, \tau)) ,  \cb( \zeta(\by, \tau)) \right] &= c^2(\tau) \int \md^3 x \md^3 y \left[3 \zeta'(\bx,\tau) \zeta^2(\bx,\tau), \zeta^3(\by,\tau) \right] \nonumber\\
&= \frac{-i c^2(\tau)}{2 \mpl^2 a^2(\tau) \epsilon(\tau)} \int \md^3 x ~9\zeta^4(\bx,\tau) 
\end{align}
Then, contribution from cubic Hamiltonian to the one-loop correction reads
\begin{align}
\langle \mathcal{O}(\tau_0) \rangle^{(3)} = \int \md^3x \int_{-\infty}^{\tau_0} \mathrm{d}\tau ~\mathrm{Re} \left\langle \mathcal{\hat{O}} (\tau_0) \frac{9i c^2(\tau)}{2 \mpl^2 a^2(\tau) \epsilon(\tau)} \zeta^4 (\bx,\tau) \right\rangle,
\end{align}
From \eqref{4ind}, the quartic-induced Hamiltonian reads
\begin{equation}
H_\cb^\mathrm{(4I)}(\tau) = \int \md^3 x \frac{9 c^2(\tau)}{2 \mpl^2 a^2(\tau) \epsilon(\tau)} \zeta^4(\bx,\tau),
\end{equation}
and it contributes
\begin{align}
\langle \mathcal{O}(\tau_0) \rangle^\mathrm{(4I)} &= 2 \int_{-\infty}^{\tau_0} \mathrm{d}\tau ~\mathrm{Im} \left\langle \mathcal{\hat{O}} (\tau_0) H_\cb^\mathrm{(4I)}(\tau) \right\rangle \nonumber\\
&= \int \md^3x \int_{-\infty}^{\tau_0} \mathrm{d}\tau ~\mathrm{Im} \left\langle \mathcal{\hat{O}} (\tau_0) \frac{9 c^2(\tau)}{2 \mpl^2 a^2(\tau) \epsilon(\tau)} \zeta^4(\bx,\tau) \right\rangle.
\end{align}
We can see that both contributions cancel with each other,
\begin{equation}
\langle \mathcal{O}(\tau_0) \rangle^{(3)} + \langle \mathcal{O}(\tau_0) \rangle^\mathrm{(4I)} = 0.
\end{equation}
Therefore, total time derivative cubic self-interaction contributes nothing to correlation functions of $\zeta$.

\section{Source method \label{app:source}}
As an alternative, the one-loop correction to the large-scale power spectrum can be derived by solving the equation of motion with interaction term \eqref{eom}, which is called the source method \cite{Kristiano:2023scm,Riotto:2023gpm}. For a superhorizon perturbation, the third term in \eqref{eom} can be neglected, so the equation of motion becomes
\begin{equation}
\zeta_\bp'' + \frac{(a^2 \epsilon)'}{a^2 \epsilon} \zeta_\bp' = - \frac{(a^2 \epsilon \eta')'}{4a^2 \epsilon} \int \frac{\md^3 k}{(2\pi)^3} \zeta_\bk \zeta_{\bp-\bk} .
\end{equation}
The inhomogeneous solution reads
\begin{equation}
\zeta^s_\bp(\tau) = - \frac{1}{4} \int_{-\infty}^{\tau} \frac{\md \tau_1}{a^2(\tau_1) \epsilon(\tau_1)} \int_{-\infty}^{\tau_1} \md \tau_2 \left[ a^2(\tau_2) \epsilon(\tau_2) \eta'(\tau_2) \right]' \int \frac{\md^3 k}{(2 \pi)^3} \zeta_\bk(\tau_2) \zeta_{\bp-\bk}(\tau_2),
\end{equation}
then performing integration by parts leads to
\begin{align}
\zeta^s_\bp(\tau_0) = & - \frac{1}{4} \int_{-\infty}^{\tau_0} \frac{\md \tau_1}{a^2(\tau_1) \epsilon(\tau_1)} a^2(\tau_1) \epsilon(\tau_1) \eta'(\tau_1) \int \frac{\md^3 k}{(2 \pi)^3} \zeta_\bk(\tau_1) \zeta_{\bp-\bk}(\tau_1) \\
& + \frac{1}{4} \int_{-\infty}^{\tau_0} \frac{\md \tau_1}{a^2(\tau_1) \epsilon(\tau_1)} \int_{-\infty}^{\tau_1} \md \tau_2 ~ a^2(\tau_2) \epsilon(\tau_2) \eta'(\tau_2) \int \frac{\md^3 k}{(2 \pi)^3} \frac{\md}{\md \tau_2} \left[ \zeta_\bk(\tau_2) \zeta_{\bp-\bk}(\tau_2) \right]. \nonumber
\end{align}
Evaluating the integral with $\eta'(\tau) = \Delta\eta ~\delta(\tau-\tau_e)$ yields
\begin{equation}
\zeta^s_\bp(\tau_0) = - \frac{1}{4} \Delta\eta \int \frac{\md^3 k}{(2 \pi)^3} \zeta_\bk(\tau_e) \zeta_{\bp-\bk}(\tau_e) + \frac{1}{4} \Delta\eta \int \frac{\md^3 k}{(2 \pi)^3} \frac{2}{3 k_e} \zeta'_\bk(\tau_e) \zeta_{\bp-\bk}(\tau_e). 
\end{equation}

Here is the subtle point, time derivative of $\zeta_\bk(\tau)$ satisfies Heisenberg equation of motion
\begin{equation}
\frac{\md}{\md\tau} \zeta_\bk(\tau) = i [H^{(3)}(\tau) , \zeta_\bk(\tau)] + \frac{\partial}{\partial\tau} \zeta_\bk(\tau).
\end{equation}
Substituting \eqref{hint} to the commutator yields
\begin{align}
i [H^{(3)}(\tau) , \zeta_\bk(\tau)] = - \frac{i}{2} \mpl^2 a^2(\tau) \epsilon(\tau) \eta'(\tau) \int & \frac{\md^3k_1 \md^3k_2 \md^3k_3}{(2 \pi)^9} (2 \pi)^3 \delta(\bk_1 + \bk_2 + \bk_3) \nonumber\\
& \times [\zeta'_{\bk_1}(\tau) \zeta_{\bk_2}(\tau) \zeta_{\bk_3}(\tau), \zeta_{\bk}(\tau) ].
\end{align}
Evaluating the commutator leads to
\begin{align}
i [H^{(3)}(\tau) , \zeta_\bk(\tau)] & = - \frac{i}{2} \mpl^2 a^2(\tau) \epsilon(\tau) \eta'(\tau) (-2i) \mathrm{Im}[\zeta_k(\tau) \zeta_k'^*(\tau)] \int \frac{\md^3 q}{(2\pi)^3} \zeta_\bq(\tau) \zeta_{\bk-\bq}(\tau) \nonumber\\
& = - \frac{1}{4} \eta'(\tau) \int \frac{\md^3 q}{(2\pi)^3} \zeta_\bq(\tau) \zeta_{\bk-\bq}(\tau).
\end{align}
Finally, correlating the inhomogeneous solution with homogeneous (free) solution yields
\begin{align}
\langle \! \langle \zeta^s_{\bp}(\tau_0) \zeta^f_{-\bp}(\tau_0) \rangle \! \rangle = \frac{1}{4} \Delta\eta \int \frac{\md^3 k}{(2 \pi)^3} & \left[ - \langle \! \langle \zeta_\bk(\tau_e) \zeta_{\bp-\bk}(\tau_e) \zeta_{-\bp}(\tau_e) \rangle \! \rangle + \frac{2}{3 k_e} \langle \! \langle \zeta'_\bk(\tau_e) \zeta_{\bp-\bk}(\tau_e) \zeta_{-\bp}(\tau_e) \rangle \! \rangle  \right. \nonumber\\
& \left. + \frac{2}{3 k_e} \langle \! \langle i [H^{(3)}(\tau_e) , \zeta_\bk(\tau_e)] \zeta_{\bp-\bk}(\tau_e) \zeta_{-\bp}(\tau_e) \rangle \! \rangle \right],
\end{align}
where the first two terms are the one-loop correction generated by cubic Hamiltonian \eqref{loopsharp} and last term is precisely the one-loop correction generated by quartic-induced Hamiltonian \eqref{4indloop}\footnote{Indeed, we did not incorporate this term in \cite{Kristiano:2023scm}.}
\begin{equation}
\langle \! \langle \zeta_{\bp}(\tau_0) \zeta_{-\bp}(\tau_0) \rangle \! \rangle^\mathrm{(4I)} = - \frac{\Delta\eta }{4 k_e}  \eta'(\tau_e) |\zeta_p(\tau_0)|^2 \int \frac{\md^3 k}{(2 \pi)^3}  |\zeta_k(\tau_e)|^2 .
\end{equation}
Therefore, the one-loop correction from quartic-induced Hamiltonian can be derived by source method.

\bibliographystyle{jhep}
\bibliography{Reference}

\providecommand{\noopsort}[1]{}\providecommand{\singleletter}[1]{#1}%

\providecommand{\href}[2]{#2}\begingroup\raggedright\begin{thebibliography}{100}

\bibitem{Kristiano:2022maq}
J.~Kristiano and J.~Yokoyama, \emph{{Constraining Primordial Black Hole Formation from Single-Field Inflation}}, \href{https://doi.org/10.1103/PhysRevLett.132.221003}{\emph{Phys. Rev. Lett.} {\bfseries 132} (2024) 221003} [\href{https://arxiv.org/abs/2211.03395}{{\ttfamily 2211.03395}}].

\bibitem{Kristiano:2023scm}
J.~Kristiano and J.~Yokoyama, \emph{{Note on the bispectrum and one-loop corrections in single-field inflation with primordial black hole formation}}, \href{https://doi.org/10.1103/PhysRevD.109.103541}{\emph{Phys. Rev. D} {\bfseries 109} (2024) 103541} [\href{https://arxiv.org/abs/2303.00341}{{\ttfamily 2303.00341}}].

\bibitem{Kristiano:2024ngc}
J.~Kristiano and J.~Yokoyama, \emph{{Generating large primordial fluctuations in single-field inflation for PBH formation}},  \href{https://arxiv.org/abs/2405.12149}{{\ttfamily 2405.12149}}.

\bibitem{Kristiano:2021urj}
J.~Kristiano and J.~Yokoyama, \emph{{Why Must Primordial Non-Gaussianity Be Very Small?}}, \href{https://doi.org/10.1103/PhysRevLett.128.061301}{\emph{Phys. Rev. Lett.} {\bfseries 128} (2022) 061301} [\href{https://arxiv.org/abs/2104.01953}{{\ttfamily 2104.01953}}].

\bibitem{Kristiano:2022zpn}
J.~Kristiano and J.~Yokoyama, \emph{{Perturbative region on non-Gaussian parameter space in single-field inflation}}, \href{https://doi.org/10.1088/1475-7516/2022/07/007}{\emph{JCAP} {\bfseries 07} (2022) 007} [\href{https://arxiv.org/abs/2204.05202}{{\ttfamily 2204.05202}}].

\bibitem{Riotto:2023hoz}
A.~Riotto, \emph{{The Primordial Black Hole Formation from Single-Field Inflation is Not Ruled Out}},  \href{https://arxiv.org/abs/2301.00599}{{\ttfamily 2301.00599}}.

\bibitem{Riotto:2023gpm}
A.~Riotto, \emph{{The Primordial Black Hole Formation from Single-Field Inflation is Still Not Ruled Out}},  \href{https://arxiv.org/abs/2303.01727}{{\ttfamily 2303.01727}}.

\bibitem{Choudhury:2023vuj}
S.~Choudhury, M.R.~Gangopadhyay and M.~Sami, \emph{{No-go for the formation of heavy mass Primordial Black Holes in Single Field Inflation}},  \href{https://arxiv.org/abs/2301.10000}{{\ttfamily 2301.10000}}.

\bibitem{Choudhury:2023jlt}
S.~Choudhury, S.~Panda and M.~Sami, \emph{{PBH formation in EFT of single field inflation with sharp transition}}, \href{https://doi.org/10.1016/j.physletb.2023.138123}{\emph{Phys. Lett. B} {\bfseries 845} (2023) 138123} [\href{https://arxiv.org/abs/2302.05655}{{\ttfamily 2302.05655}}].

\bibitem{Choudhury:2023rks}
S.~Choudhury, S.~Panda and M.~Sami, \emph{{Quantum loop effects on the power spectrum and constraints on primordial black holes}}, \href{https://doi.org/10.1088/1475-7516/2023/11/066}{\emph{JCAP} {\bfseries 11} (2023) 066} [\href{https://arxiv.org/abs/2303.06066}{{\ttfamily 2303.06066}}].

\bibitem{Franciolini:2023lgy}
G.~Franciolini, A.~Iovino, Junior., M.~Taoso and A.~Urbano, \emph{{One loop to rule them all: Perturbativity in the presence of ultra slow-roll dynamics}},  \href{https://arxiv.org/abs/2305.03491}{{\ttfamily 2305.03491}}.

\bibitem{Davies:2023hhn}
M.W.~Davies, L.~Iacconi and D.J.~Mulryne, \emph{{Numerical 1-loop correction from a potential yielding ultra-slow-roll dynamics}}, \href{https://doi.org/10.1088/1475-7516/2024/04/050}{\emph{JCAP} {\bfseries 04} (2024) 050} [\href{https://arxiv.org/abs/2312.05694}{{\ttfamily 2312.05694}}].

\bibitem{Jackson:2023obv}
J.H.P.~Jackson, H.~Assadullahi, A.D.~Gow, K.~Koyama, V.~Vennin and D.~Wands, \emph{{The separate-universe approach and sudden transitions during inflation}},  \href{https://arxiv.org/abs/2311.03281}{{\ttfamily 2311.03281}}.

\bibitem{Mishra:2023lhe}
S.S.~Mishra, E.J.~Copeland and A.M.~Green, \emph{{Primordial black holes and stochastic inflation beyond slow roll. Part I. Noise matrix elements}}, \href{https://doi.org/10.1088/1475-7516/2023/09/005}{\emph{JCAP} {\bfseries 09} (2023) 005} [\href{https://arxiv.org/abs/2303.17375}{{\ttfamily 2303.17375}}].

\bibitem{Choudhury:2024jlz}
S.~Choudhury, A.~Karde, P.~Padiyar and M.~Sami, \emph{{Primordial Black Holes from Effective Field Theory of Stochastic Single Field Inflation at NNNLO}},  \href{https://arxiv.org/abs/2403.13484}{{\ttfamily 2403.13484}}.

\bibitem{Firouzjahi:2023ahg}
H.~Firouzjahi and A.~Riotto, \emph{{Primordial Black Holes and loops in single-field inflation}}, \href{https://doi.org/10.1088/1475-7516/2024/02/021}{\emph{JCAP} {\bfseries 02} (2024) 021} [\href{https://arxiv.org/abs/2304.07801}{{\ttfamily 2304.07801}}].

\bibitem{Iacconi:2023ggt}
L.~Iacconi, D.~Mulryne and D.~Seery, \emph{{Loop corrections in the separate universe picture}},  \href{https://arxiv.org/abs/2312.12424}{{\ttfamily 2312.12424}}.

\bibitem{Motohashi:2023syh}
H.~Motohashi and Y.~Tada, \emph{{Squeezed bispectrum and one-loop corrections in transient constant-roll inflation}}, \href{https://doi.org/10.1088/1475-7516/2023/08/069}{\emph{JCAP} {\bfseries 08} (2023) 069} [\href{https://arxiv.org/abs/2303.16035}{{\ttfamily 2303.16035}}].

\bibitem{Tasinato:2023ukp}
G.~Tasinato, \emph{{Large |\ensuremath{\eta}| approach to single field inflation}}, \href{https://doi.org/10.1103/PhysRevD.108.043526}{\emph{Phys. Rev. D} {\bfseries 108} (2023) 043526} [\href{https://arxiv.org/abs/2305.11568}{{\ttfamily 2305.11568}}].

\bibitem{Tasinato:2023ioq}
G.~Tasinato, \emph{{Non-Gaussianities and the large |\ensuremath{\eta}| approach to inflation}}, \href{https://doi.org/10.1103/PhysRevD.109.063510}{\emph{Phys. Rev. D} {\bfseries 109} (2024) 063510} [\href{https://arxiv.org/abs/2312.03498}{{\ttfamily 2312.03498}}].

\bibitem{Firouzjahi:2024psd}
H.~Firouzjahi, \emph{{Loop Corrections in Bispectrum in USR Inflation with PBHs Formation}},  \href{https://arxiv.org/abs/2403.03841}{{\ttfamily 2403.03841}}.

\bibitem{Maity:2023qzw}
S.~Maity, H.V.~Ragavendra, S.K.~Sethi and L.~Sriramkumar, \emph{{Loop contributions to the scalar power spectrum due to quartic order action in ultra slow roll inflation}},  \href{https://arxiv.org/abs/2307.13636}{{\ttfamily 2307.13636}}.

\bibitem{Firouzjahi:2023aum}
H.~Firouzjahi, \emph{{One-loop corrections in power spectrum in single field inflation}}, \href{https://doi.org/10.1088/1475-7516/2023/10/006}{\emph{JCAP} {\bfseries 10} (2023) 006} [\href{https://arxiv.org/abs/2303.12025}{{\ttfamily 2303.12025}}].

\bibitem{Fumagalli:2023hpa}
J.~Fumagalli, \emph{{Absence of one-loop effects on large scales from small scales in non-slow-roll dynamics}},  \href{https://arxiv.org/abs/2305.19263}{{\ttfamily 2305.19263}}.

\bibitem{Tada:2023rgp}
Y.~Tada, T.~Terada and J.~Tokuda, \emph{{Cancellation of quantum corrections on the soft curvature perturbations}}, \href{https://doi.org/10.1007/JHEP01(2024)105}{\emph{JHEP} {\bfseries 01} (2024) 105} [\href{https://arxiv.org/abs/2308.04732}{{\ttfamily 2308.04732}}].

\bibitem{Firouzjahi:2023bkt}
H.~Firouzjahi, \emph{{Revisiting loop corrections in single field ultraslow-roll inflation}}, \href{https://doi.org/10.1103/PhysRevD.109.043514}{\emph{Phys. Rev. D} {\bfseries 109} (2024) 043514} [\href{https://arxiv.org/abs/2311.04080}{{\ttfamily 2311.04080}}].

\bibitem{Cheng:2023ikq}
S.-L.~Cheng, D.-S.~Lee and K.-W.~Ng, \emph{{Primordial perturbations from ultra-slow-roll single-field inflation with quantum loop effects}}, \href{https://doi.org/10.1088/1475-7516/2024/03/008}{\emph{JCAP} {\bfseries 03} (2024) 008} [\href{https://arxiv.org/abs/2305.16810}{{\ttfamily 2305.16810}}].

\bibitem{Saburov:2024und}
S.~Saburov and S.V.~Ketov, \emph{{Quantum loop corrections in the modified gravity model of Starobinsky inflation with primordial black hole production}},  \href{https://arxiv.org/abs/2402.02934}{{\ttfamily 2402.02934}}.

\bibitem{Ballesteros:2024zdp}
G.~Ballesteros and J.G.~Egea, \emph{{One-loop power spectrum in ultra slow-roll inflation and implications for primordial black hole dark matter}},  \href{https://arxiv.org/abs/2404.07196}{{\ttfamily 2404.07196}}.

\bibitem{Inomata:2024lud}
K.~Inomata, \emph{{Curvature Perturbations Protected Against One Loop}},  \href{https://arxiv.org/abs/2403.04682}{{\ttfamily 2403.04682}}.

\bibitem{Carr:2009jm}
B.J.~Carr, K.~Kohri, Y.~Sendouda and J.~Yokoyama, \emph{{New cosmological constraints on primordial black holes}}, \href{https://doi.org/10.1103/PhysRevD.81.104019}{\emph{Phys. Rev. D} {\bfseries 81} (2010) 104019} [\href{https://arxiv.org/abs/0912.5297}{{\ttfamily 0912.5297}}].

\bibitem{Carr:2020gox}
B.~Carr, K.~Kohri, Y.~Sendouda and J.~Yokoyama, \emph{{Constraints on primordial black holes}}, \href{https://doi.org/10.1088/1361-6633/ac1e31}{\emph{Rept. Prog. Phys.} {\bfseries 84} (2021) 116902} [\href{https://arxiv.org/abs/2002.12778}{{\ttfamily 2002.12778}}].

\bibitem{Starobinsky:1980te}
A.A.~Starobinsky, \emph{{A New Type of Isotropic Cosmological Models Without Singularity}}, \href{https://doi.org/10.1016/0370-2693(80)90670-X}{\emph{Phys. Lett. B} {\bfseries 91} (1980) 99}.

\bibitem{Sato:1980yn}
K.~Sato, \emph{{First Order Phase Transition of a Vacuum and Expansion of the Universe}}, {\emph{Mon. Not. Roy. Astron. Soc.} {\bfseries 195} (1981) 467}.

\bibitem{Guth:1980zm}
A.H.~Guth, \emph{{The Inflationary Universe: A Possible Solution to the Horizon and Flatness Problems}}, \href{https://doi.org/10.1103/PhysRevD.23.347}{\emph{Phys. Rev. D} {\bfseries 23} (1981) 347}.

\bibitem{Sato:2015dga}
K.~Sato and J.~Yokoyama, \emph{{Inflationary cosmology: First 30+ years}}, \href{https://doi.org/10.1142/S0218271815300256}{\emph{Int. J. Mod. Phys. D} {\bfseries 24} (2015) 1530025}.

\bibitem{Planck:2018nkj}
{\scshape Planck} collaboration, \emph{{Planck 2018 results. I. Overview and the cosmological legacy of Planck}}, \href{https://doi.org/10.1051/0004-6361/201833880}{\emph{Astron. Astrophys.} {\bfseries 641} (2020) A1} [\href{https://arxiv.org/abs/1807.06205}{{\ttfamily 1807.06205}}].

\bibitem{Planck:2018jri}
{\scshape Planck} collaboration, \emph{{Planck 2018 results. X. Constraints on inflation}}, \href{https://doi.org/10.1051/0004-6361/201833887}{\emph{Astron. Astrophys.} {\bfseries 641} (2020) A10} [\href{https://arxiv.org/abs/1807.06211}{{\ttfamily 1807.06211}}].

\bibitem{Planck:2019kim}
{\scshape Planck} collaboration, \emph{{Planck 2018 results. IX. Constraints on primordial non-Gaussianity}}, \href{https://doi.org/10.1051/0004-6361/201935891}{\emph{Astron. Astrophys.} {\bfseries 641} (2020) A9} [\href{https://arxiv.org/abs/1905.05697}{{\ttfamily 1905.05697}}].

\bibitem{Nakama:2014vla}
T.~Nakama, T.~Suyama and J.~Yokoyama, \emph{{Reheating the Universe Once More: The Dissipation of Acoustic Waves as a Novel Probe of Primordial Inhomogeneities on Even Smaller Scales}}, \href{https://doi.org/10.1103/PhysRevLett.113.061302}{\emph{Phys. Rev. Lett.} {\bfseries 113} (2014) 061302} [\href{https://arxiv.org/abs/1403.5407}{{\ttfamily 1403.5407}}].

\bibitem{Jeong:2014gna}
D.~Jeong, J.~Pradler, J.~Chluba and M.~Kamionkowski, \emph{{Silk damping at a redshift of a billion: a new limit on small-scale adiabatic perturbations}}, \href{https://doi.org/10.1103/PhysRevLett.113.061301}{\emph{Phys. Rev. Lett.} {\bfseries 113} (2014) 061301} [\href{https://arxiv.org/abs/1403.3697}{{\ttfamily 1403.3697}}].

\bibitem{Inomata:2016uip}
K.~Inomata, M.~Kawasaki and Y.~Tada, \emph{{Revisiting constraints on small scale perturbations from big-bang nucleosynthesis}}, \href{https://doi.org/10.1103/PhysRevD.94.043527}{\emph{Phys. Rev. D} {\bfseries 94} (2016) 043527} [\href{https://arxiv.org/abs/1605.04646}{{\ttfamily 1605.04646}}].

\bibitem{Nakama:2017qac}
T.~Nakama, T.~Suyama, K.~Kohri and N.~Hiroshima, \emph{{Constraints on small-scale primordial power by annihilation signals from extragalactic dark matter minihalos}}, \href{https://doi.org/10.1103/PhysRevD.97.023539}{\emph{Phys. Rev. D} {\bfseries 97} (2018) 023539} [\href{https://arxiv.org/abs/1712.08820}{{\ttfamily 1712.08820}}].

\bibitem{Kawasaki:2021yek}
M.~Kawasaki, H.~Nakatsuka and K.~Nakayama, \emph{{Constraints on small-scale primordial density fluctuation from cosmic microwave background through dark matter annihilation}}, \href{https://doi.org/10.1088/1475-7516/2022/03/061}{\emph{JCAP} {\bfseries 03} (2022) 061} [\href{https://arxiv.org/abs/2110.12620}{{\ttfamily 2110.12620}}].

\bibitem{Kimura:2021sqz}
R.~Kimura, T.~Suyama, M.~Yamaguchi and Y.-L.~Zhang, \emph{{Reconstruction of Primordial Power Spectrum of curvature perturbation from the merger rate of Primordial Black Hole Binaries}}, \href{https://doi.org/10.1088/1475-7516/2021/04/031}{\emph{JCAP} {\bfseries 04} (2021) 031} [\href{https://arxiv.org/abs/2102.05280}{{\ttfamily 2102.05280}}].

\bibitem{Wang:2022nml}
X.~Wang, Y.-l.~Zhang, R.~Kimura and M.~Yamaguchi, \emph{{Reconstruction of power spectrum of primordial curvature perturbations on small scales from primordial black hole binaries scenario of LIGO/VIRGO detection}}, \href{https://doi.org/10.1007/s11433-023-2091-x}{\emph{Sci. China Phys. Mech. Astron.} {\bfseries 66} (2023) 260462} [\href{https://arxiv.org/abs/2209.12911}{{\ttfamily 2209.12911}}].

\bibitem{Ivanov:1994pa}
P.~Ivanov, P.~Naselsky and I.~Novikov, \emph{{Inflation and primordial black holes as dark matter}}, \href{https://doi.org/10.1103/PhysRevD.50.7173}{\emph{Phys. Rev. D} {\bfseries 50} (1994) 7173}.

\bibitem{Kinney:1997ne}
W.H.~Kinney, \emph{{A Hamilton-Jacobi approach to nonslow roll inflation}}, \href{https://doi.org/10.1103/PhysRevD.56.2002}{\emph{Phys. Rev. D} {\bfseries 56} (1997) 2002} [\href{https://arxiv.org/abs/hep-ph/9702427}{{\ttfamily hep-ph/9702427}}].

\bibitem{Inoue:2001zt}
J.~Yokoyama and S.~Inoue, \emph{{Curvature perturbation at the local extremum of the inflaton's potential}}, \href{https://doi.org/10.1016/S0370-2693(01)01369-7}{\emph{Phys. Lett. B} {\bfseries 524} (2002) 15} [\href{https://arxiv.org/abs/hep-ph/0104083}{{\ttfamily hep-ph/0104083}}].

\bibitem{Kinney:2005vj}
W.H.~Kinney, \emph{{Horizon crossing and inflation with large eta}}, \href{https://doi.org/10.1103/PhysRevD.72.023515}{\emph{Phys. Rev. D} {\bfseries 72} (2005) 023515} [\href{https://arxiv.org/abs/gr-qc/0503017}{{\ttfamily gr-qc/0503017}}].

\bibitem{Martin:2012pe}
J.~Martin, H.~Motohashi and T.~Suyama, \emph{{Ultra Slow-Roll Inflation and the non-Gaussianity Consistency Relation}}, \href{https://doi.org/10.1103/PhysRevD.87.023514}{\emph{Phys. Rev. D} {\bfseries 87} (2013) 023514} [\href{https://arxiv.org/abs/1211.0083}{{\ttfamily 1211.0083}}].

\bibitem{Motohashi:2017kbs}
H.~Motohashi and W.~Hu, \emph{{Primordial Black Holes and Slow-Roll Violation}}, \href{https://doi.org/10.1103/PhysRevD.96.063503}{\emph{Phys. Rev. D} {\bfseries 96} (2017) 063503} [\href{https://arxiv.org/abs/1706.06784}{{\ttfamily 1706.06784}}].

\bibitem{Motohashi:2014ppa}
H.~Motohashi, A.A.~Starobinsky and J.~Yokoyama, \emph{{Inflation with a constant rate of roll}}, \href{https://doi.org/10.1088/1475-7516/2015/09/018}{\emph{JCAP} {\bfseries 09} (2015) 018} [\href{https://arxiv.org/abs/1411.5021}{{\ttfamily 1411.5021}}].

\bibitem{Motohashi:2017aob}
H.~Motohashi and A.A.~Starobinsky, \emph{{Constant-roll inflation: confrontation with recent observational data}}, \href{https://doi.org/10.1209/0295-5075/117/39001}{\emph{EPL} {\bfseries 117} (2017) 39001} [\href{https://arxiv.org/abs/1702.05847}{{\ttfamily 1702.05847}}].

\bibitem{Motohashi:2019rhu}
H.~Motohashi, S.~Mukohyama and M.~Oliosi, \emph{{Constant Roll and Primordial Black Holes}}, \href{https://doi.org/10.1088/1475-7516/2020/03/002}{\emph{JCAP} {\bfseries 03} (2020) 002} [\href{https://arxiv.org/abs/1910.13235}{{\ttfamily 1910.13235}}].

\bibitem{Yokoyama:1998rw}
J.~Yokoyama, \emph{{Chaotic new inflation and primordial spectrum of adiabatic fluctuations}}, \href{https://doi.org/10.1103/PhysRevD.59.107303}{\emph{Phys. Rev. D} {\bfseries 59} (1999) 107303}.

\bibitem{Saito:2008em}
R.~Saito, J.~Yokoyama and R.~Nagata, \emph{{Single-field inflation, anomalous enhancement of superhorizon fluctuations, and non-Gaussianity in primordial black hole formation}}, \href{https://doi.org/10.1088/1475-7516/2008/06/024}{\emph{JCAP} {\bfseries 06} (2008) 024} [\href{https://arxiv.org/abs/0804.3470}{{\ttfamily 0804.3470}}].

\bibitem{Zel:1967}
Y.B.~{Zel'dovich} and I.D.~{Novikov}, \emph{{The Hypothesis of Cores Retarded during Expansion and the Hot Cosmological Model}}, {\emph{Sov. Astron.} {\bfseries 10} (1967) 602}.

\bibitem{Hawking:1971ei}
S.~Hawking, \emph{{Gravitationally collapsed objects of very low mass}}, {\emph{Mon. Not. Roy. Astron. Soc.} {\bfseries 152} (1971) 75}.

\bibitem{Carr:1974nx}
B.J.~Carr and S.W.~Hawking, \emph{{Black holes in the early Universe}}, {\emph{Mon. Not. Roy. Astron. Soc.} {\bfseries 168} (1974) 399}.

\bibitem{Hawking:1974rv}
S.W.~Hawking, \emph{{Black hole explosions}}, \href{https://doi.org/10.1038/248030a0}{\emph{Nature} {\bfseries 248} (1974) 30}.

\bibitem{Chapline:1975ojl}
G.F.~Chapline, \emph{{Cosmological effects of primordial black holes}}, \href{https://doi.org/10.1038/253251a0}{\emph{Nature} {\bfseries 253} (1975) 251}.

\bibitem{Garcia-Bellido:1996mdl}
J.~Garcia-Bellido, A.D.~Linde and D.~Wands, \emph{{Density perturbations and black hole formation in hybrid inflation}}, \href{https://doi.org/10.1103/PhysRevD.54.6040}{\emph{Phys. Rev. D} {\bfseries 54} (1996) 6040} [\href{https://arxiv.org/abs/astro-ph/9605094}{{\ttfamily astro-ph/9605094}}].

\bibitem{Yokoyama:1995ex}
J.~Yokoyama, \emph{{Formation of MACHO primordial black holes in inflationary cosmology}}, {\emph{Astron. Astrophys.} {\bfseries 318} (1997) 673} [\href{https://arxiv.org/abs/astro-ph/9509027}{{\ttfamily astro-ph/9509027}}].

\bibitem{Afshordi:2003zb}
N.~Afshordi, P.~McDonald and D.N.~Spergel, \emph{{Primordial black holes as dark matter: The Power spectrum and evaporation of early structures}}, \href{https://doi.org/10.1086/378763}{\emph{Astrophys. J. Lett.} {\bfseries 594} (2003) L71} [\href{https://arxiv.org/abs/astro-ph/0302035}{{\ttfamily astro-ph/0302035}}].

\bibitem{Frampton:2010sw}
P.H.~Frampton, M.~Kawasaki, F.~Takahashi and T.T.~Yanagida, \emph{{Primordial Black Holes as All Dark Matter}}, \href{https://doi.org/10.1088/1475-7516/2010/04/023}{\emph{JCAP} {\bfseries 04} (2010) 023} [\href{https://arxiv.org/abs/1001.2308}{{\ttfamily 1001.2308}}].

\bibitem{Belotsky:2014kca}
K.M.~Belotsky, A.D.~Dmitriev, E.A.~Esipova, V.A.~Gani, A.V.~Grobov, M.Y.~Khlopov et~al., \emph{{Signatures of primordial black hole dark matter}}, \href{https://doi.org/10.1142/S0217732314400057}{\emph{Mod. Phys. Lett. A} {\bfseries 29} (2014) 1440005} [\href{https://arxiv.org/abs/1410.0203}{{\ttfamily 1410.0203}}].

\bibitem{Carr:2016drx}
B.~Carr, F.~Kuhnel and M.~Sandstad, \emph{{Primordial Black Holes as Dark Matter}}, \href{https://doi.org/10.1103/PhysRevD.94.083504}{\emph{Phys. Rev. D} {\bfseries 94} (2016) 083504} [\href{https://arxiv.org/abs/1607.06077}{{\ttfamily 1607.06077}}].

\bibitem{Inomata:2017okj}
K.~Inomata, M.~Kawasaki, K.~Mukaida, Y.~Tada and T.T.~Yanagida, \emph{{Inflationary Primordial Black Holes as All Dark Matter}}, \href{https://doi.org/10.1103/PhysRevD.96.043504}{\emph{Phys. Rev. D} {\bfseries 96} (2017) 043504} [\href{https://arxiv.org/abs/1701.02544}{{\ttfamily 1701.02544}}].

\bibitem{Espinosa:2017sgp}
J.R.~Espinosa, D.~Racco and A.~Riotto, \emph{{Cosmological Signature of the Standard Model Higgs Vacuum Instability: Primordial Black Holes as Dark Matter}}, \href{https://doi.org/10.1103/PhysRevLett.120.121301}{\emph{Phys. Rev. Lett.} {\bfseries 120} (2018) 121301} [\href{https://arxiv.org/abs/1710.11196}{{\ttfamily 1710.11196}}].

\bibitem{Green:2020jor}
A.M.~Green and B.J.~Kavanagh, \emph{{Primordial Black Holes as a dark matter candidate}}, \href{https://doi.org/10.1088/1361-6471/abc534}{\emph{J. Phys. G} {\bfseries 48} (2021) 043001} [\href{https://arxiv.org/abs/2007.10722}{{\ttfamily 2007.10722}}].

\bibitem{Carr:2020xqk}
B.~Carr and F.~Kuhnel, \emph{{Primordial Black Holes as Dark Matter: Recent Developments}}, \href{https://doi.org/10.1146/annurev-nucl-050520-125911}{\emph{Ann. Rev. Nucl. Part. Sci.} {\bfseries 70} (2020) 355} [\href{https://arxiv.org/abs/2006.02838}{{\ttfamily 2006.02838}}].

\bibitem{Sasaki:2016jop}
M.~Sasaki, T.~Suyama, T.~Tanaka and S.~Yokoyama, \emph{{Primordial Black Hole Scenario for the Gravitational-Wave Event GW150914}}, \href{https://doi.org/10.1103/PhysRevLett.117.061101}{\emph{Phys. Rev. Lett.} {\bfseries 117} (2016) 061101} [\href{https://arxiv.org/abs/1603.08338}{{\ttfamily 1603.08338}}].

\bibitem{NANOGrav:2023gor}
{\scshape NANOGrav} collaboration, \emph{{The NANOGrav 15 yr Data Set: Evidence for a Gravitational-wave Background}}, \href{https://doi.org/10.3847/2041-8213/acdac6}{\emph{Astrophys. J. Lett.} {\bfseries 951} (2023) L8} [\href{https://arxiv.org/abs/2306.16213}{{\ttfamily 2306.16213}}].

\bibitem{NANOGrav:2023hvm}
{\scshape NANOGrav} collaboration, \emph{{The NANOGrav 15 yr Data Set: Search for Signals from New Physics}}, \href{https://doi.org/10.3847/2041-8213/acdc91}{\emph{Astrophys. J. Lett.} {\bfseries 951} (2023) L11} [\href{https://arxiv.org/abs/2306.16219}{{\ttfamily 2306.16219}}].

\bibitem{Fumagalli:2023loc}
J.~Fumagalli, S.~Bhattacharya, M.~Peloso, S.~Renaux-Petel and L.T.~Witkowski, \emph{{One-loop infrared rescattering by enhanced scalar fluctuations during inflation}}, \href{https://doi.org/10.1088/1475-7516/2024/04/029}{\emph{JCAP} {\bfseries 04} (2024) 029} [\href{https://arxiv.org/abs/2307.08358}{{\ttfamily 2307.08358}}].

\bibitem{Inomata:2022yte}
K.~Inomata, M.~Braglia, X.~Chen and S.~Renaux-Petel, \emph{{Questions on calculation of primordial power spectrum with large spikes: the resonance model case}}, \href{https://doi.org/10.1088/1475-7516/2023/04/011}{\emph{JCAP} {\bfseries 04} (2023) 011} [\href{https://arxiv.org/abs/2211.02586}{{\ttfamily 2211.02586}}].

\bibitem{Caravano:2024tlp}
A.~Caravano, K.~Inomata and S.~Renaux-Petel, \emph{{The Inflationary Butterfly Effect: Non-Perturbative Dynamics From Small-Scale Features}},  \href{https://arxiv.org/abs/2403.12811}{{\ttfamily 2403.12811}}.

\bibitem{Cai:2018dkf}
Y.-F.~Cai, X.~Chen, M.H.~Namjoo, M.~Sasaki, D.-G.~Wang and Z.~Wang, \emph{{Revisiting non-Gaussianity from non-attractor inflation models}}, \href{https://doi.org/10.1088/1475-7516/2018/05/012}{\emph{JCAP} {\bfseries 05} (2018) 012} [\href{https://arxiv.org/abs/1712.09998}{{\ttfamily 1712.09998}}].

\bibitem{Sou:2022nsd}
C.M.~Sou, D.H.~Tran and Y.~Wang, \emph{{Decoherence of cosmological perturbations from boundary terms and the non-classicality of gravity}}, \href{https://doi.org/10.1007/JHEP04(2023)092}{\emph{JHEP} {\bfseries 04} (2023) 092} [\href{https://arxiv.org/abs/2207.04435}{{\ttfamily 2207.04435}}].

\bibitem{Ning:2023ybc}
S.~Ning, C.M.~Sou and Y.~Wang, \emph{{On the decoherence of primordial gravitons}}, \href{https://doi.org/10.1007/JHEP06(2023)101}{\emph{JHEP} {\bfseries 06} (2023) 101} [\href{https://arxiv.org/abs/2305.08071}{{\ttfamily 2305.08071}}].

\bibitem{Braglia:2024zsl}
M.~Braglia and L.~Pinol, \emph{{No time to derive: unraveling total time derivatives in in-in perturbation theory}},  \href{https://arxiv.org/abs/2403.14558}{{\ttfamily 2403.14558}}.

\bibitem{Kawaguchi:2024lsw}
R.~Kawaguchi, S.~Tsujikawa and Y.~Yamada, \emph{{Roles of boundary and equation-of-motion terms in cosmological correlation functions}},  \href{https://arxiv.org/abs/2403.16022}{{\ttfamily 2403.16022}}.

\bibitem{Syu:2019uwx}
W.-C.~Syu, D.-S.~Lee and K.-W.~Ng, \emph{{Quantum loop effects to the power spectrum of primordial perturbations during ultra slow-roll inflation}}, \href{https://doi.org/10.1103/PhysRevD.101.025013}{\emph{Phys. Rev. D} {\bfseries 101} (2020) 025013} [\href{https://arxiv.org/abs/1907.13089}{{\ttfamily 1907.13089}}].

\bibitem{Cheng:2021lif}
S.-L.~Cheng, D.-S.~Lee and K.-W.~Ng, \emph{{Power spectrum of primordial perturbations during ultra-slow-roll inflation with back reaction effects}}, \href{https://doi.org/10.1016/j.physletb.2022.136956}{\emph{Phys. Lett. B} {\bfseries 827} (2022) 136956} [\href{https://arxiv.org/abs/2106.09275}{{\ttfamily 2106.09275}}].

\bibitem{Starobinsky:1992ts}
A.A.~Starobinsky, \emph{{Spectrum of adiabatic perturbations in the universe when there are singularities in the inflation potential}}, {\emph{JETP Lett.} {\bfseries 55} (1992) 489}.

\bibitem{Leach:2001zf}
S.M.~Leach, M.~Sasaki, D.~Wands and A.R.~Liddle, \emph{{Enhancement of superhorizon scale inflationary curvature perturbations}}, \href{https://doi.org/10.1103/PhysRevD.64.023512}{\emph{Phys. Rev. D} {\bfseries 64} (2001) 023512} [\href{https://arxiv.org/abs/astro-ph/0101406}{{\ttfamily astro-ph/0101406}}].

\bibitem{Byrnes:2018txb}
C.T.~Byrnes, P.S.~Cole and S.P.~Patil, \emph{{Steepest growth of the power spectrum and primordial black holes}}, \href{https://doi.org/10.1088/1475-7516/2019/06/028}{\emph{JCAP} {\bfseries 06} (2019) 028} [\href{https://arxiv.org/abs/1811.11158}{{\ttfamily 1811.11158}}].

\bibitem{Liu:2020oqe}
J.~Liu, Z.-K.~Guo and R.-G.~Cai, \emph{{Analytical approximation of the scalar spectrum in the ultraslow-roll inflationary models}}, \href{https://doi.org/10.1103/PhysRevD.101.083535}{\emph{Phys. Rev. D} {\bfseries 101} (2020) 083535} [\href{https://arxiv.org/abs/2003.02075}{{\ttfamily 2003.02075}}].

\bibitem{Tasinato:2020vdk}
G.~Tasinato, \emph{{An analytic approach to non-slow-roll inflation}}, \href{https://doi.org/10.1103/PhysRevD.103.023535}{\emph{Phys. Rev. D} {\bfseries 103} (2021) 023535} [\href{https://arxiv.org/abs/2012.02518}{{\ttfamily 2012.02518}}].

\bibitem{Karam:2022nym}
A.~Karam, N.~Koivunen, E.~Tomberg, V.~Vaskonen and H.~Veerm\"ae, \emph{{Anatomy of single-field inflationary models for primordial black holes}}, \href{https://doi.org/10.1088/1475-7516/2023/03/013}{\emph{JCAP} {\bfseries 03} (2023) 013} [\href{https://arxiv.org/abs/2205.13540}{{\ttfamily 2205.13540}}].

\bibitem{Pi:2022zxs}
S.~Pi and J.~Wang, \emph{{Primordial black hole formation in Starobinsky's linear potential model}}, \href{https://doi.org/10.1088/1475-7516/2023/06/018}{\emph{JCAP} {\bfseries 06} (2023) 018} [\href{https://arxiv.org/abs/2209.14183}{{\ttfamily 2209.14183}}].

\bibitem{Wands:1998yp}
D.~Wands, \emph{{Duality invariance of cosmological perturbation spectra}}, \href{https://doi.org/10.1103/PhysRevD.60.023507}{\emph{Phys. Rev. D} {\bfseries 60} (1999) 023507} [\href{https://arxiv.org/abs/gr-qc/9809062}{{\ttfamily gr-qc/9809062}}].

\bibitem{Maldacena:2002vr}
J.M.~Maldacena, \emph{{Non-Gaussian features of primordial fluctuations in single field inflationary models}}, \href{https://doi.org/10.1088/1126-6708/2003/05/013}{\emph{JHEP} {\bfseries 05} (2003) 013} [\href{https://arxiv.org/abs/astro-ph/0210603}{{\ttfamily astro-ph/0210603}}].

\bibitem{Arroja:2011yj}
F.~Arroja and T.~Tanaka, \emph{{A note on the role of the boundary terms for the non-Gaussianity in general k-inflation}}, \href{https://doi.org/10.1088/1475-7516/2011/05/005}{\emph{JCAP} {\bfseries 05} (2011) 005} [\href{https://arxiv.org/abs/1103.1102}{{\ttfamily 1103.1102}}].

\bibitem{Burrage:2011hd}
C.~Burrage, R.H.~Ribeiro and D.~Seery, \emph{{Large slow-roll corrections to the bispectrum of noncanonical inflation}}, \href{https://doi.org/10.1088/1475-7516/2011/07/032}{\emph{JCAP} {\bfseries 07} (2011) 032} [\href{https://arxiv.org/abs/1103.4126}{{\ttfamily 1103.4126}}].

\bibitem{Namjoo:2012aa}
M.H.~Namjoo, H.~Firouzjahi and M.~Sasaki, \emph{{Violation of non-Gaussianity consistency relation in a single field inflationary model}}, \href{https://doi.org/10.1209/0295-5075/101/39001}{\emph{EPL} {\bfseries 101} (2013) 39001} [\href{https://arxiv.org/abs/1210.3692}{{\ttfamily 1210.3692}}].

\bibitem{Cai:2016ngx}
Y.-F.~Cai, J.-O.~Gong, D.-G.~Wang and Z.~Wang, \emph{{Features from the non-attractor beginning of inflation}}, \href{https://doi.org/10.1088/1475-7516/2016/10/017}{\emph{JCAP} {\bfseries 10} (2016) 017} [\href{https://arxiv.org/abs/1607.07872}{{\ttfamily 1607.07872}}].

\bibitem{Chen:2013eea}
X.~Chen, H.~Firouzjahi, E.~Komatsu, M.H.~Namjoo and M.~Sasaki, \emph{{In-in and $\delta N$ calculations of the bispectrum from non-attractor single-field inflation}}, \href{https://doi.org/10.1088/1475-7516/2013/12/039}{\emph{JCAP} {\bfseries 12} (2013) 039} [\href{https://arxiv.org/abs/1308.5341}{{\ttfamily 1308.5341}}].

\bibitem{Passaglia:2018ixg}
S.~Passaglia, W.~Hu and H.~Motohashi, \emph{{Primordial black holes and local non-Gaussianity in canonical inflation}}, \href{https://doi.org/10.1103/PhysRevD.99.043536}{\emph{Phys. Rev. D} {\bfseries 99} (2019) 043536} [\href{https://arxiv.org/abs/1812.08243}{{\ttfamily 1812.08243}}].

\bibitem{Davies:2021loj}
M.W.~Davies, P.~Carrilho and D.J.~Mulryne, \emph{{Non-Gaussianity in inflationary scenarios for primordial black holes}}, \href{https://doi.org/10.1088/1475-7516/2022/06/019}{\emph{JCAP} {\bfseries 06} (2022) 019} [\href{https://arxiv.org/abs/2110.08189}{{\ttfamily 2110.08189}}].

\bibitem{Chen:2009bc}
X.~Chen, B.~Hu, M.-x.~Huang, G.~Shiu and Y.~Wang, \emph{{Large Primordial Trispectra in General Single Field Inflation}}, \href{https://doi.org/10.1088/1475-7516/2009/08/008}{\emph{JCAP} {\bfseries 08} (2009) 008} [\href{https://arxiv.org/abs/0905.3494}{{\ttfamily 0905.3494}}].

\bibitem{Akhshik:2015nfa}
M.~Akhshik, H.~Firouzjahi and S.~Jazayeri, \emph{{Effective Field Theory of non-Attractor Inflation}}, \href{https://doi.org/10.1088/1475-7516/2015/07/048}{\emph{JCAP} {\bfseries 07} (2015) 048} [\href{https://arxiv.org/abs/1501.01099}{{\ttfamily 1501.01099}}].

\bibitem{Arkani-Hamed:2018kmz}
N.~Arkani-Hamed, D.~Baumann, H.~Lee and G.L.~Pimentel, \emph{{The Cosmological Bootstrap: Inflationary Correlators from Symmetries and Singularities}}, \href{https://doi.org/10.1007/JHEP04(2020)105}{\emph{JHEP} {\bfseries 04} (2020) 105} [\href{https://arxiv.org/abs/1811.00024}{{\ttfamily 1811.00024}}].

\bibitem{Pajer:2020wxk}
E.~Pajer, \emph{{Building a Boostless Bootstrap for the Bispectrum}}, \href{https://doi.org/10.1088/1475-7516/2021/01/023}{\emph{JCAP} {\bfseries 01} (2021) 023} [\href{https://arxiv.org/abs/2010.12818}{{\ttfamily 2010.12818}}].

\end{thebibliography}\endgroup

\end{document}